\documentclass[useAMS,usenatbib]{mn2e}
\usepackage{graphicx}

\usepackage{aas_macros}
\usepackage{amssymb}
\usepackage{amsmath}

\newcommand{\bri}{\hbox{$B\!RI$}}
\newcommand{\gvri}{\hbox{$gV\!RI$}}
\newcommand{\bvri}{\hbox{$BV\!RI$}}

\newcommand{\about}{$\sim\!\!$~}
\newcommand{\ri}{\hbox{$R-I$}}
\newcommand{\bmax}{\hbox{$B_{\rm max}$}}
\newcommand{\be}{\begin{displaymath}}
\newcommand{\ee}{\end{displaymath}}

\def\lsim{\hbox{\rlap{\raise 0.425ex\hbox{$<$}}\lower 0.65ex\hbox{$\sim$}}}
\def\gsim{\hbox{\rlap{\raise 0.425ex\hbox{$>$}}\lower 0.65ex\hbox{$\sim$}}}
\def\arcmin{\hbox{$^\prime$}}
\def\arcsec{\hbox{$^{\prime\prime}$}}

\newcommand{\msun}{M$_\odot$}

\newcommand{\halpha}{H$\alpha$}

\newcommand{\hbeta}{H$\beta$}

\newcommand{\kms}{km~s$^{-1}$}
\newcommand{\perMpc}{Mpc$^{-1}$}
\newcommand{\ergps}{erg~s$^{-1}$}

\newcommand{\nic}{$^{56}$Ni}

\newcommand{\ion}[2]{#1$\;${\small{#2}}\relax}
\newcommand{\bv}{\hbox{$B-V$}}
\newcommand{\vr}{\hbox{$V-R$}}

\newcommand{\dc}{SN~2009dc}
\newcommand{\gz}{SN~2006gz}
\newcommand{\fg}{SN~2003fg}
\newcommand{\snif}{SN~2007if}

\title[SN~2009dc, Super-Chandrasekhar Mass?]{Fourteen Months of
  Observations of the Possible Super-Chandrasekhar Mass Type Ia
  Supernova 2009dc} 

\author[Silverman, et~al.]{Jeffrey M. Silverman,$^{1,2}$\thanks{E-mail:
    JSilverman@astro.berkeley.edu} Mohan Ganeshalingam,$^{1}$ Weidong
  Li,$^{1}$
\newauthor
Alexei V. Filippenko,$^{1}$ Adam A. Miller,$^{1}$ and Dovi Poznanski$^{1}$\\
$^{1}$Department of Astronomy, University of California,
Berkeley, CA 94720-3411, USA\\
$^{2}$Marc J. Staley Fellow
}

\begin{document}

\date{Accepted  . Received   ; in original form  }

\pagerange{\pageref{firstpage}--\pageref{lastpage}} \pubyear{2010}

\maketitle

\label{firstpage}

\begin{abstract}
In this paper, we present and analyse optical photometry and spectra
of the extremely luminous and slowly evolving Type~Ia supernova
(SN~Ia) 2009dc, and offer evidence that it is a super-Chandrasekhar
mass (SC) SN~Ia and thus had a SC white dwarf (WD) progenitor. Optical
spectra of SN~2007if, a similar object, are also shown. \dc\ had one
of the most slowly evolving light curves ever observed for a SN~Ia,
with a rise time of \about23~days and $\Delta m_{15} (B)=0.72$~mag. We
calculate a lower limit to the peak bolometric luminosity of
\about$2.4\times10^{43}$~\ergps, though the actual value is likely
almost 40\% larger.  Optical spectra of \dc\ and \snif\ obtained near
maximum brightness exhibit strong \ion{C}{II} features (indicative of
a significant amount of unburned material), and the post-maximum
spectra are dominated by iron-group elements.  All of our spectra of
\dc\ and \snif\ also show low expansion velocities. However, we see no
strong evidence in \dc\ for a velocity ``plateau'' near maximum light
like the one seen in \snif\ \citep{Scalzo10}. The high
luminosity and low expansion velocities of \dc\ lead us to derive a
possible WD progenitor mass of more than 2~\msun\ and a \nic\ mass of
about 1.4--1.7~\msun.  We propose that the host galaxy of
\dc\ underwent a gravitational interaction with a neighboring galaxy
in the relatively recent past. This may have led to a sudden burst of
star formation which could have produced the SC~WD progenitor of
\dc\ and likely turned the neighboring galaxy into a ``post-starburst
galaxy.''  No published model seems to match the extreme values
observed in \dc, but simulations do show that such massive progenitors
can exist (likely as a result of the merger of two WDs) and can
possibly explode as SC~SNe~Ia.

\end{abstract}



\section{Introduction}\label{s:intro}

Type~Ia supernovae (SNe~Ia) are differentiated from other types of SNe
by the absence of hydrogen and the presence of broad absorption from
\ion{Si}{II} $\lambda$6355 in their optical spectra \citep[for a
  review see][]{Filippenko97}.  SNe~Ia have been used to measure
cosmological parameters to high precision
\citep[e.g.,][]{Kowalski08,Hicken09,Kessler09,Amanullah10}, as well as
to discover the accelerating expansion of the Universe
\citep{Riess98,Perlmutter99}.  Broadly speaking, SNe~Ia are the result
of thermonuclear explosions of C-O white dwarfs (WDs) resulting from
either the accretion of matter from a nondegenerate companion star
\citep[e.g.,][]{Whelan73} or the merger of two degenerate objects
\citep[e.g.,][]{Iben84,Webbink84}.  However, the nature of the
companion and the details of the explosion itself are both still quite
uncertain.

The cosmological utility of SNe~Ia comes from the fact that they 
are standardizable candles (i.e., their luminosities at peak can be
calibrated).  This naively seems reasonable since SNe~Ia should all
have the same amount of fuel and the same trigger point: they should
all explode when a WD nearly reaches the Chandrasekhar mass of
\about1.4~\msun.   

In 2003, SNLS-03D3bb (also known as \fg) was discovered
\citep{Howell06} and was shown to be a SN~Ia that was overluminous by
about a factor of 2, and had a slowly declining light curve, quite low
expansion velocities, and unburned material present in near-maximum
light spectra.  This last observation implies that a layer of carbon
and oxygen from the progenitor existed on top of the burned silicon
layer.  In SNe~Ia, carbon is usually extremely weak or completely
absent (even at very early times) in optical and infrared (IR)
spectra, although it has been (sometimes tentatively) identified in a
few other cases \citep[e.g.,][]{Branch03,Marion06,Thomas07,Foley10}.
\citet{Howell06} suggest that all of the oddities seen in \fg\ could
be explained if its progenitor WD had a mass greater then the
canonical upper limit for WDs of \about1.4~\msun, a so-called
``super-Chandrasekhar mass'' (SC) WD.

\citet{Hicken07} then presented data on \gz\ which shared some of the
strange properties of \fg, and they concluded that \gz\ must have come
from a WD merger leading to a SC~SN~Ia.  Recently, \citet{Scalzo10}
published observations of \snif\ which is yet another example of this
emerging class of possible SC~SNe~Ia.  They not only observed low
expansion velocities, but they also saw a plateau in the expansion
velocity near maximum-brightness which they interpret as evidence for
the SN ejecta running into a shell of material.  The accurately
determined total mass of the \snif\ system is well above the 
Chandrasekhar mass.

\begin{figure}
\centering
\includegraphics[width=8.5cm]{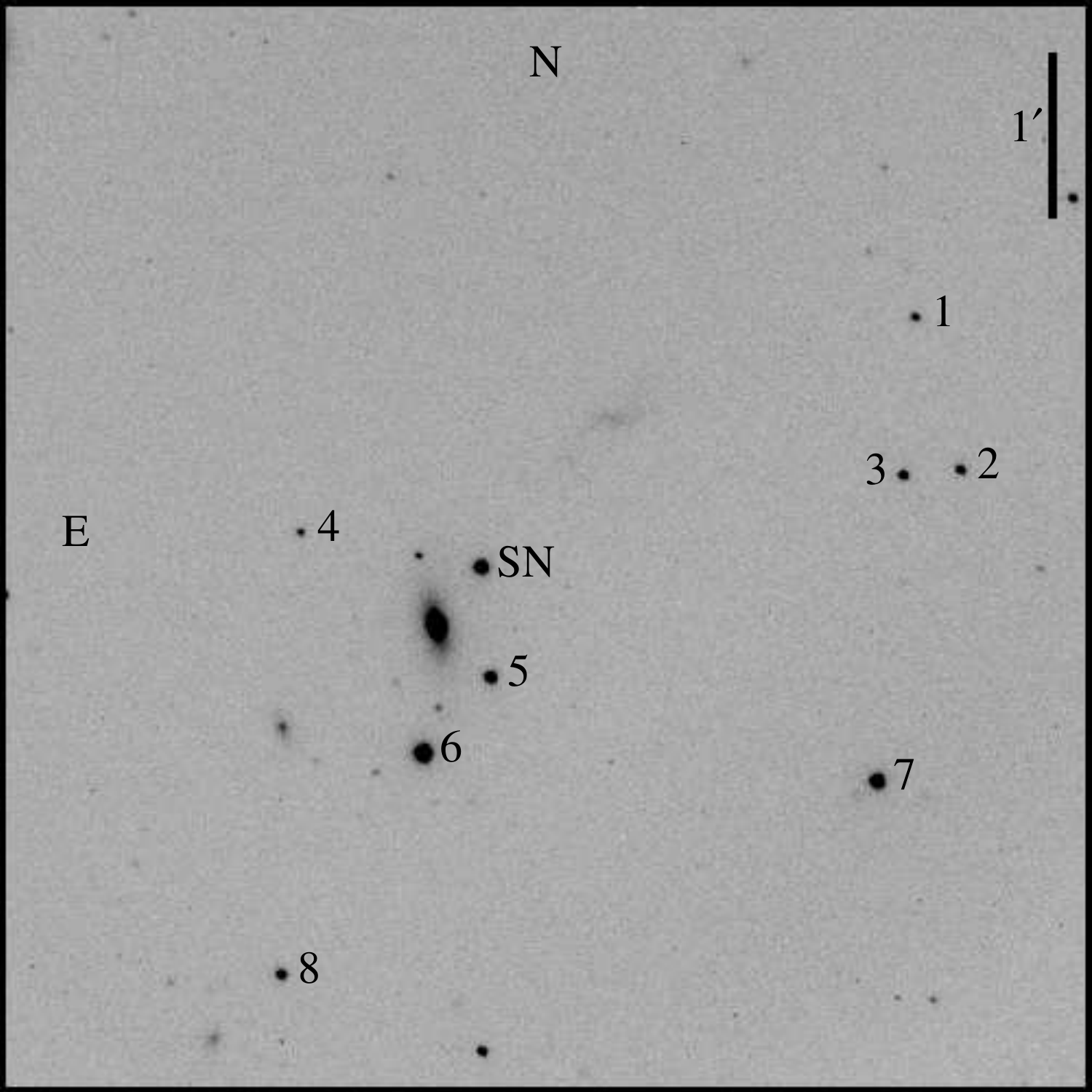}
\caption{KAIT image of \dc\ and its host galaxy, UGC~10064. The field
 of view is 6.7\arcmin$\times$6.7\arcmin. \dc\ is labeled along with the 
 comparison stars used for differential photometry. The scale of the
 image is marked in the top right; north is up and east is to left. 
 The SN is sufficiently far from its host galaxy that template subtraction 
 was not performed when conducting photometry.}\label{f:galaxy}   
\end{figure}

Recently, it was pointed out that there might be another member of
this SC~SN~Ia class,
\dc\ \citep{Harutyunyan09,Marion09,Yamanaka09}. \dc\ was discovered
$15\farcs8$ west and $20\farcs8$ north of the nucleus of the S0 galaxy
UGC~10064 by \citet{Puckett09} on 2009~Apr.~9.31 (UT dates are used
throughout this paper), though in \S\ref{ss:lc} we will show a
detection of the SN \about5~days earlier.  It is located at
$\alpha_{\rm J2000} = 15^{\rm h}51^{\rm m}12\fs12$ and $\delta_{\rm
  J2000} = +25\degr42\arcmin28\farcs0$; the SN and its host are shown
in Figure~\ref{f:galaxy}.  No object was visible in our data at the
position of the SN on 2009~Mar.~28 to a limiting magnitude of
\about19.5, so the actual explosion date was almost certainly between
Mar.~28 and Apr.~4.

\citet{Harutyunyan09} obtained a spectrum of \dc\ one week after the
announced discovery date and showed it to be a SN~Ia before maximum
light. They also noted that \dc\ spectroscopically resembled the
SC~SN~Ia candidate SN~2006gz \citep{Hicken07} at this time, but with a
much lower expansion velocity as derived from the \ion{Si}{II}
$\lambda$6355 absorption feature.  Three days later, nearly
simultaneous optical and IR spectra were obtained by \citet{Marion09},
covering a wavelength range of 0.36 to 1.3 $\mu$m.  They again note
some similarities to (as well as differences from) SN~2006gz, as well
as similarities with the possible SC~SN~Ia 2003fg \citep{Howell06}.

\citet{Yamanaka09} presented early-time optical and near-IR
observations of \dc\ and showed that it did indeed share many of the
properties seen in both \fg\ and \gz, suggesting that it too is a
possible SC~SN~Ia.  Furthermore, \citet{Tanaka09} published
spectropolarimetry of \dc\ which indicated that the explosion was
quite spherically symmetric.

In this paper we present and analyse our own optical photometric and
spectroscopic data for \dc\ (as well as spectra of \snif) with the
goal of more definitively answering the question of whether \dc\ and
\snif\ were truly SC~SNe~Ia. We show some of the most precise data on
the rise time of a possible SC~SN~Ia ever published, as well as some
of the latest-time photometric and spectral observations of a member
of this class.  In \S\ref{s:obs} we describe our observations and data
reduction, and in \S\ref{s:analysis} we discuss our analysis of the
photometry and spectra of \dc\ (and its host galaxy) and \snif. We
attempt to robustly calculate physical parameters of \dc\ using a
variety of methods and compare them to theoretical predictions in
\S\ref{s:discussion}. Finally, in \S\ref{s:conclusions}, we summarize
our conclusions and ruminate about the future.


\section{Observations and Data Reduction}\label{s:obs}

\subsection{Photometry}

Observations of \dc\ began on 2009~Apr.~17, about one week before
maximum $B$-band brightness, in \bvri\ filters using the 0.76-m
Katzman Automatic Imaging Telescope \citep[KAIT;][]{Filippenko01} and
the 1-m Nickel telescope, both at Lick Observatory. We continued to
follow \dc\ for over 5~months until 2009~Sep.~26, when it reached the
western limit of both telescopes in the early evening.  We obtained
late-time \gvri\ images of \dc\ using the dual-arm Low-Resolution
Imaging Spectrometer \citep[LRIS;][]{Oke95} with the 10-m Keck~I
telescope on 6~Feb.~2010 (281~days past maximum), \bri\ images using
the DEIMOS spectrograph \citep{Faber03} mounted on the 10-m Keck~II
telescope on 12~June~2010 (403~days past maximum), and $V$-band images
again using LRIS on 13~June~2010.

Our optical photometry is complemented with data taken from the
UltraViolet Optical Telescope (UVOT) on the {\it Swift} Observatory in
the $U$, $B$, $V$, $UVW1$, $UVM2$, and $UVW2$ filters. We downloaded 7
epochs of observations from the {\it Swift} archives.

Our early-time optical images were reduced using a pipeline developed
for KAIT and Nickel data \citep{Ganeshalingam10:phot_paper}. The
images were bias subtracted and flatfielded at the telescope. Given
the distance of \dc\ from the host galaxy, we did not find it
necessary to attempt galaxy subtraction. We performed differential
photometry using point-spread function (PSF) fitting photometry on
\dc\ and several comparison stars in the field with the DAOPHOT
package in IRAF.\footnote{IRAF: The Image Reduction and Analysis
  Facility is distributed by the National Optical Astronomy
  Observatory, which is operated by the Association of Universities
  for Research in Astronomy (AURA) under cooperative agreement with
  the National Science Foundation (NSF).} The output instrumental
magnitudes were calibrated to the standard Johnson $BV$ and Cousins
$RI$ system using colour terms derived from many photometric
nights. The comparison stars were calibrated against \citet{Landolt92}
standard stars using 13 photometric epochs to achieve a root-mean
square (rms) of $<0.01$~mag.  The comparison stars are identified in
Figure~\ref{f:galaxy} with corresponding photometry in
Table~\ref{comp_stars}.  To estimate the uncertainty in our photometry
pipeline, we add artificial stars with the same magnitude and PSF as
the SN to each data image at positions of similar background
brightness; the rms of doing this procedure 20 times is taken as the
uncertainty. The photometric uncertainty is added in quadrature to the
calibration error to produce our final uncertainty, adopting an error
floor of 0.03~mag for $B$ and $I$ and 0.02~mag for $V$ and $R$ to
account for slight systematic differences between the KAIT and Nickel
photometry. Our final photometry of \dc\ is presented in
Table~\ref{t:phot_table}.

\begin{table*}
 \caption{Comparison Stars\label{comp_stars} for the \dc\ Field}
\begin{center}
\begin{minipage}{14cm}
\begin{center}
\begin{tabular*}{14cm}{lccccccc}
\hline
Star & $\alpha_{\rm J2000}$ &$\delta_{\rm J2000}$ &$B$ (mag)& $V$ (mag) & $R$ (mag) &
$I$ (mag) & $N_\textrm{calib}$ \\
\hline
 SN & 15:51:12.12 & +25:42:28.0 & & & & & \\
  1 & 15:51:00.40 & +25:44:00.1 &   18.769 (010) &   17.418 (006) &   16.503 (005) &   15.744 (011) &  9\\
  2 & 15:50:59.16 & +25:43:04.7 &   17.501 (008) &   16.957 (006) &   16.608 (004) &   16.236 (008) & 12\\
  3 & 15:51:00.72 & +25:43:02.5 &   18.126 (009) &   16.816 (004) &   15.940 (004) &   15.228 (008) & 10\\
  4 & 15:51:16.96 & +25:42:40.6 &   18.671 (006) &   17.805 (005) &   17.249 (006) &   16.734 (011) & 11\\
  5 & 15:51:11.83 & +25:41:48.3 &   16.661 (007) &   15.872 (004) &   15.406 (003) &   14.984 (009) & 13\\
  6 & 15:51:13.64 & +25:41:20.5 &   15.383 (007) &   14.471 (004) &   13.935 (003) &   13.460 (009) & 13\\
  7 & 15:51:01.38 & +25:41:11.2 &   15.892 (007) &   15.200 (005) &   14.808 (003) &   14.420 (010) & 13\\
  8 & 15:51:17.44 & +25:39:59.6 &   17.156 (006) &   16.534 (004) &   16.174 (004) &   15.790 (008) & 13\\
\hline
\end{tabular*}
\end{center}
$1\sigma$ uncertainties are in parentheses, in units of 0.001~mag.
\end{minipage}
\end{center}
\end{table*}

 \begin{table*}
 \caption{Early-time \bvri\ photometry of \dc\  \label{t:phot_table}}
\begin{center}
\begin{minipage}{11.5cm}
\begin{center}
\begin{tabular*}{11.5cm}{lccccc}
\hline
JD & $B$ (mag) & $V$ (mag)& $R$ (mag)& $I$ (mag) &Telescope \\
\hline
2454925.95 & $\cdots$      & $\cdots$      & 18.824  (177) & $\cdots$      & KAIT \\
2454938.98 & 15.670  (030) & 15.638  (020) & 15.596  (020) & 15.718  (030) & KAIT \\
2454939.95 & 15.549  (030) & 15.541  (020) & 15.500  (020) & 15.577  (030) & Nickel\\
2454940.92 & 15.474  (030) & 15.472  (030) & 15.430  (030) & 15.524  (030) & Nickel\\
2454940.93 & 15.511  (030) & 15.495  (020) & 15.475  (020) & 15.608  (030) & KAIT \\
2454942.95 & 15.455  (030) & 15.409  (020) & 15.376  (020) & 15.529  (030) & KAIT \\
2454946.97 & 15.383  (030) & 15.327  (020) & 15.272  (020) & 15.450  (030) & KAIT \\
2454948.87 & 15.369  (030) & 15.350  (020) & 15.288  (020) & 15.454  (030) & KAIT \\
2454951.92 & 15.449  (030) & 15.386  (020) & 15.311  (020) & 15.449  (030) & KAIT \\
2454957.97 & 15.770  (030) & 15.535  (020) & 15.423  (020) & 15.537  (030) & KAIT \\
2454959.85 & 15.930  (030) & 15.610  (020) & 15.464  (020) & 15.554  (030) & KAIT \\
2454960.92 & 16.004  (030) & 15.591  (020) & 15.476  (020) & 15.593  (030) & KAIT \\
2454962.93 & 16.137  (036) & 15.656  (024) & 15.543  (026) & 15.633  (084) & KAIT \\
2454964.90 & 16.322  (030) & 15.762  (020) & 15.584  (020) & 15.568  (030) & KAIT \\
2454966.93 & 16.528  (030) & 15.846  (020) & 15.642  (020) & 15.562  (030) & KAIT \\
2454966.95 & 16.478  (030) & 15.839  (020) & 15.611  (020) & 15.481  (030) & Nickel\\
2454968.89 & 16.677  (030) & 15.926  (020) & 15.675  (020) & 15.565  (030) & KAIT \\
2454970.88 & $\cdots$      & 16.029  (086) & 15.696  (053) & $\cdots$      & KAIT \\
2454971.92 & 16.880  (030) & 16.054  (020) & 15.701  (024) & 15.500  (030) & Nickel\\
2454972.89 & 16.972  (030) & 16.106  (020) & 15.748  (020) & 15.579  (030) & KAIT \\
2454974.88 & 17.154  (030) & 16.202  (020) & 15.799  (020) & 15.557  (030) & KAIT \\
2454975.87 & 17.188  (030) & 16.239  (020) & 15.797  (020) & 15.526  (030) & Nickel\\
2454977.90 & 17.384  (030) & 16.334  (020) & 15.874  (020) & 15.659  (030) & KAIT \\
2454981.89 & 17.650  (047) & 16.488  (020) & 16.009  (020) & 15.707  (030) & KAIT \\
2454989.83 & 18.045  (078) & 16.893  (027) & 16.364  (020) & 15.979  (034) & KAIT \\
2454989.87 & 17.891  (062) & 16.855  (030) & 16.333  (030) & 15.951  (030) & Nickel\\
2454993.85 & 18.128  (080) & 16.943  (041) & 16.480  (021) & 16.167  (030) & KAIT \\
2454993.90 & 18.032  (042) & 16.960  (022) & 16.483  (020) & 16.093  (030) & Nickel\\
2454999.81 & 18.188  (057) & 17.083  (026) & 16.696  (020) & $\cdots$      & KAIT \\
2454999.83 & 18.153  (037) & 17.102  (020) & 16.675  (027) & 16.315  (030) & Nickel\\
2455004.78 & 18.208  (033) & 17.180  (033) & 16.835  (028) & 16.478  (034) & Nickel\\
2455007.84 & 18.265  (030) & 17.289  (022) & 16.913  (027) & 16.540  (031) & Nickel\\
2455009.80 & 18.254  (089) & 17.277  (029) & 16.944  (022) & 16.677  (039) & KAIT \\
2455014.83 & 18.449  (084) & 17.344  (057) & 17.151  (021) & 16.761  (038) & KAIT \\
2455015.83 & 18.370  (046) & 17.427  (022) & 17.132  (031) & 16.794  (042) & Nickel\\
2455019.85 & 18.377  (106) & 17.432  (043) & 17.247  (040) & 16.937  (036) & Nickel\\
2455022.74 & 18.271  (097) & 17.581  (047) & 17.295  (025) & 16.890  (050) & KAIT \\
2455025.83 & 18.585  (049) & 17.607  (069) & 17.431  (059) & 17.176  (053) & Nickel\\
2455027.69 & 18.502  (054) & 17.625  (041) & 17.439  (035) & 17.243  (036) & KAIT \\
2455032.69 & 18.512  (089) & 17.770  (043) & 17.570  (020) & 17.343  (036) & KAIT \\
2455032.85 & 18.607  (032) & 17.726  (020) & 17.588  (025) & 17.317  (043) & Nickel\\
2455034.73 & 18.678  (030) & 17.779  (036) & 17.643  (028) & 17.425  (046) & Nickel\\
2455037.69 & 18.686  (053) & 17.805  (045) & 17.725  (023) & 17.455  (035) & KAIT \\
2455040.77 & 18.756  (030) & 17.902  (023) & 17.800  (021) & 17.504  (030) & Nickel\\
2455042.68 & 18.646  (062) & 18.060  (077) & 17.908  (066) & 17.690  (111) & KAIT \\
2455042.77 & 18.844  (060) & 17.912  (037) & 17.886  (035) & 17.586  (037) & Nickel\\
2455044.73 & 18.788  (041) & 17.971  (043) & 17.946  (028) & 17.666  (053) & Nickel\\
2455047.68 & 19.035  (151) & 17.965  (055) & 18.007  (116) & 17.646  (057) & KAIT \\
2455047.73 & 18.900  (142) & 18.075  (076) & 18.026  (049) & 17.825  (119) & Nickel\\
2455052.68 & 18.992  (115) & 18.208  (050) & 18.178  (040) & 17.860  (140) & KAIT \\
2455054.77 & $\cdots$      & 18.231  (083) & 18.256  (132) & 17.909  (133) & Nickel\\
2455057.69 & 18.967  (174) & 18.233  (136) & 18.198  (170) & 18.064  (176) & KAIT \\
2455059.71 & 19.026  (048) & 18.241  (036) & 18.309  (041) & 18.065  (044) & Nickel\\
2455062.67 & 19.087  (120) & 18.323  (049) & 18.364  (135) & 18.227  (174) & KAIT \\
2455064.70 & 19.070  (051) & 18.339  (045) & 18.484  (038) & 18.093  (050) & Nickel\\
2455067.66 & 19.061  (086) & 18.324  (067) & 18.462  (068) & 18.307  (172) & KAIT \\
2455068.69 & 19.144  (045) & 18.368  (024) & 18.559  (065) & 18.285  (080) & Nickel\\
2455071.69 & 19.166  (052) & 18.427  (033) & 18.632  (039) & 18.283  (090) & Nickel\\
2455072.66 & 19.164  (142) & 18.512  (075) & 18.653  (085) & 18.437  (130) & KAIT \\
2455074.69 & 19.249  (098) & 18.497  (044) & 18.717  (061) & 18.313  (082) & Nickel\\
2455077.65 & $\cdots$      & 18.725  (082) & $\cdots$      & $\cdots$      & KAIT \\
2455082.65 & 19.163  (089) & 18.789  (088) & 19.011  (188) & 18.618  (143) & KAIT \\
2455087.64 & $\cdots$      & 18.919  (171) & $\cdots$      & $\cdots$      & KAIT \\
2455090.68 & 19.444  (052) & 18.765  (037) & 19.219  (091) & 18.800  (081) & Nickel\\
2455093.66 & 19.528  (078) & 18.820  (043) & 19.176  (086) & 18.867  (144) & Nickel\\
2455100.66 & 19.665  (061) & 18.954  (095) & 19.441  (151) & $\cdots$      & Nickel\\
\hline
\end{tabular*}
\end{center}
$1\sigma$ uncertainties are in parentheses, in units of 0.001~mag.
\end{minipage}
\end{center}
\end{table*}

Late-time data obtained at the Keck~I and Keck~II telescopes were bias
subtracted and flatfielded using standard imaging
techniques. Differential photometry was performed using PSF fitting
photometry on the SN and comparison stars that were not saturated, but
also detected in our calibration images obtained with the Nickel
telescope.  Calibrations for the $g$ band were obtained using the
transformations presented by \cite{Jester05}. In cases where all of
the field stars from our Nickel calibration were saturated,
calibrations for fainter stars in the field were obtained from the
Sloan Digital Sky Survey (SDSS) and transformed into \bvri\ using
transformations for stars from \cite{Jester05}. Colour-term
corrections were not applied. We include a systematic error of
0.03~mag in all bands. Our final late-time photometry of \dc\ is
presented in Table~\ref{t:late_time_phot}, which also includes an
epoch of photometry obtained with the Nickel.

 \begin{table*}
\caption{Late-time photometry of \dc\label{t:late_time_phot}}
\begin{center}
\begin{minipage}{11.25cm}
\begin{center}
\begin{tabular*}{11.25cm}{ l c c c c c c}
\hline
JD & Phase$^\textrm{a}$ & Telescope & Filter & Exposure (s) & Mag
& $\sigma$ \\
\hline
2455202.07 & 250 & Nickel &$B$ & 600 & 21.868  & 0.268 \\
                       & 250 & Nickel &$V$ & 360 & 21.016  & 0.320 \\ 
2455233.10 & 281 & Keck/LRIS & $g$   & 180 & 21.894  & 0.050 \\
                       & 281 & Keck/LRIS & $V$   & 180 & 21.988  & 0.042 \\
                       & 281 & Keck/LRIS & $R$   &  60  & 22.600  & 0.084 \\
                       & 281&  Keck/LRIS & $I$     & 120 & 21.483  & 0.060 \\
2455359.05 & 403 & Keck/DEIMOS &  $ B $  & 360 & 25.010 & 0.120 \\
                       & 403 & Keck/DEIMOS &  $ R $ &  450 & 24.987 &  0.143 \\
                       & 403 & Keck/DEIMOS &  $ I $   & 450 & 23.746 & 0.185 \\
2455360.10 & 404 & Keck/LRIS & V & 540 &24.834 & 0.152 \\
\hline
\end{tabular*}
\end{center}
$^\textrm{a}$Rest-frame days relative to the date of $B$-band
  maximum brightness.
\end{minipage}
\end{center}
\end{table*}

We downloaded the Level-2 UVOT data from the {\it Swift}
archive. Images taken during the same pointing were registered and
stacked to produce deeper images. We performed aperture photometry
using the recipes prescribed by \cite{Li06} for the optical data and
\cite{Poole08} for the UV data.  We modified the $U$-band data to be
in the Johnson-Cousins system using the colour corrections found in
\cite{Li06}. In general, the UVOT $B$ and $V$ photometry is in good
agreement with the photometry from our ground-based telescopes to
within $\pm0.05$~mag. The UVOT $B$-band data are systemically
brighter, which could possibly be attributed to galaxy light falling
within our aperture. Our results are given in
Table~\ref{t:uvot_table}.

 \begin{table*}
\caption{UVOT Photometry of \dc \label{t:uvot_table}}
\begin{center}
\begin{minipage}{11.75cm}
\begin{center}
\begin{tabular*}{11.75cm}{l c c c c c  l c c c}
\hline
JD & Filter & Mag & $\sigma$ & & & JD & Filter & Mag & $\sigma$ \\
\hline
 2454946.57  &  UVW2  &  17.338  &  0.044  &  &  &   2454946.57  &     $U$  &  14.720  &  0.017     \\
 2454951.85  &  UVW2  &  17.756  &  0.079  &  &  &   2454952.21  &     $U$  &  15.124  &  0.010     \\
 2454952.22  &  UVW2  &  17.987  &  0.073  &  &  &   2454955.72  &    $U$  &  15.506  &  0.022     \\
 2454955.72  &  UVW2  &  18.246  &  0.053  &  &  &   2454960.61  &     $U$  &  16.016  &  0.027     \\
 2454960.61  &  UVW2  &  18.741  &  0.066  &  &  &   2454980.35  &     $U$  &  17.890  &  0.066     \\
 2454980.35  &  UVW2  &  20.220  &  0.146  &  &   &  2454984.44  &     $U$  &  18.070  &  0.088     \\
 2454984.44  &  UVW2  &  20.314  &  0.177  &  &   &  2454946.57  &     $B$  &  15.324  &  0.017     \\
 2454946.58  &  UVM2  &  17.261  &  0.027  &  &  &   2454955.72  &     $B$  &  15.630  &  0.017     \\
 2454955.73  &  UVM2  &  18.346  &  0.058  &  &  &   2454960.61  &     $B$  &  15.935  &  0.019     \\
 2454960.61  &  UVM2  &  18.884  &  0.075  &  &  &   2454980.35  &     $B$  &  17.542  &  0.042     \\
 2454980.35  &  UVM2  &  19.986  &  0.135  &  &  &   2454984.44  &     $B$  &  17.739  &  0.055     \\
 2454984.45  &  UVM2  &  19.586  &  0.138  &  &  &   2454946.58  &     $V$  &  15.286  &  0.028     \\
 2454946.53  &  UVW1  &  16.041  &  0.032  &  &  &   2454955.73  &     $V$  &  15.489  &  0.027     \\
 2454951.71  &  UVW1  &  16.537  &  0.035  &  &   &  2454960.61  &     $V$  &  15.587  &  0.028     \\
 2454955.65  &  UVW1  &  16.965  &  0.035  &  &  &   2454980.35  &     $V$  &  16.550  &  0.044     \\
 2454960.27  &  UVW1  &  17.404  &  0.042  &  &  &   2454984.45  &     $V$  &  16.714  &  0.058     \\
 2454980.20  &  UVW1  &  18.644  &  0.077  &  &    &     &       &    &     \\
 2454984.17  &  UVW1  &  18.812  &  0.099  &  &  &       &       &    &       \\
\hline
\end{tabular*}
\end{center}
\end{minipage}
\end{center}
\end{table*}

We did not follow the photometric behaviour of \snif, another SC~SN~Ia
candidate, but we do present spectroscopic observations (see
\S\ref{ss:spec}). Throughout the rest of the paper we adopt
2007~Sep.~5.4 as the date of $B$-band maximum brightness and 0.07416
as the redshift ($z$) of \snif, both taken from \citet{Scalzo10}.

\subsection{Spectroscopy}\label{ss:spec}

Beginning about a week before maximum brightness, optical spectra of
\dc\ were obtained mainly using the dual-arm Kast spectrograph
\citep{Miller93} on the Lick 3-m Shane telescope.  Two spectra were
also obtained using LRIS on Keck~I.  Our last spectral observation
occurred 281~days after $B$-band maximum.

The Kast spectra all used a 2\arcsec\ wide slit, a 600/4310 grism on
the blue side, and a 300/7500 grating on the red side, yielding
full-width at half-maximum (FWHM) resolutions of \about4~\AA\ and
\about10~\AA, respectively.  The LRIS spectrum was obtained with a
1\arcsec\ slit, a 600/4000 grism on the blue side, and a 400/8500
grating on the red side, resulting in FWHM resolutions of
\about4~\AA\ and \about6~\AA, respectively.  All observations were
aligned along the parallactic angle to reduce differential light
losses \citep{Filippenko82}. Table~\ref{t:obs} summarizes the spectral
data of \dc\ presented in this paper.

\begin{table}
\caption{Journal of Spectroscopic Observations of \dc\label{t:obs}}
\begin{center}
\begin{tabular}{lrccc}
\hline
UT Date &Age$^\textrm{a}$&Range (\AA)&Airmass$^\textrm{b}$&Exp (s)\\
\hline
2009 Apr.\ 18.5  &$-7$ & 3500--9900\phantom{1}  & 1.06 & 1500 \\
2009 May\ 31.3  &35& 3500--10200  & 1.03 & 1500 \\
2009 Jun.\ 17.5$^\textrm{c}$ & 52  & 3400--10200  & 1.56 & 250/200$^\textrm{d}$ \\
2009 Jun.\ 29.3  &64& 3500--10200  & 1.08 & 2100 \\
2009 Jul.\ 16.3  &80& 3500--10200  & 1.29 & 2100 \\
2009 Jul.\ 23.2  &87& 3500--10200  & 1.10 & 2100 \\
2009 Jul.\ 28.3  &92& 3500--10200  & 1.33 & 2100 \\
2009 Aug.\ 14.2  &109& 3500--10000  & 1.31 & 2400 \\
2010 Feb.\ 6.6$^\textrm{c}$  &281& 3500--10200  & 1.19 & 3$\times$(630/600)$^\textrm{d}$ \\
\hline
\end{tabular}
\end{center}
$^\textrm{a}$Rest-frame days relative to the date of $B$-band
  brightness maximum, 2009~Apr.~25.4 (see \S\ref{ss:lc}).

$^\textrm{b}$Airmass at midpoint of exposure.

$^\textrm{c}$These observations used LRIS
  \citep{Oke95} on the 10-m  
Keck I telescope. The others used the Kast spectrograph on the Lick 3-m
Shane telescope \citep{Miller93}.

$^\textrm{d}$The blue side was exposed longer than the red
  side due to the relatively 
  long readout time of the red-side CCD in LRIS.
\end{table}

We obtained three spectra of \snif\ using LRIS. For the first two LRIS
spectra, we used a 400/3400 grism on the blue side and a 400/8500
grating on the red side, resulting in FWHM resolutions of
\about6~\AA\ on both sides, while the final spectrum of \snif\ was
obtained with a 600/4000 grism on the blue side, giving a resolution
of \about4 \AA. In all cases, the long, $1''$-wide slit was aligned
along the parallactic angle to reduce differential light
losses. Table~\ref{t:07if} summarizes the spectral data on
\snif\ presented here.  We also note that our last spectrum of
\snif\ (from 2007~Dec.~13) was taken under nonideal observing
conditions (clouds were present and the atmospheric seeing was poor),
and thus its spectrophotometric accuracy is not as good as that of the
other observations.

\begin{table}
\caption{Journal of Spectroscopic Observations of \snif\label{t:07if}}
\begin{center}
\begin{tabular}{lcccc}
\hline
UT Date &Age$^\textrm{a}$&Range (\AA)&Airmass$^\textrm{b}$&Exp. (s)\\
\hline
2007 Oct.\ 15.4  & 37  & 3300--9200  & 1.03 & 1200 \\
2007 Nov.\ 12.2  & 63  & 3300--9150  & 1.49 & 1200 \\
2007 Dec.\ 13.2$^\textrm{c}$  & 92  & 3300--9150  & 1.01 & 1200 \\
\hline
\end{tabular}
\end{center}
$^\textrm{a}$Rest-frame days relative to the date of $B$-band
  maximum brightness, 2007~Sep.~5.4 \citep{Scalzo10}.

$^\textrm{b}$Airmass at midpoint of exposure.

$^\textrm{c}$This observation used a slightly higher resolution 
  grism for the blue side.
\end{table}

All spectra were reduced using standard techniques
\citep[e.g.,][]{Foley03}.  Routine CCD processing and spectrum
extraction were completed with IRAF, and the data were extracted with
the optimal algorithm of \citet{Horne86}.  We obtained the wavelength
scale from low-order polynomial fits to calibration-lamp spectra.
Small wavelength shifts were then applied to the data after
cross-correlating a template sky to the night-sky lines that were
extracted with the SN.  Using our own IDL routines, we fit
spectrophotometric standard-star spectra to the data in order to flux
calibrate our spectra and to remove telluric lines \citep{Wade88,
  Matheson00}. Information regarding both our photometric and
spectroscopic data (such as observing conditions, instrument, reducer,
etc.) was obtained from our SN database (SNDB).  The SNDB uses the
popular open-source software stack known as LAMP: the Linux operating
system, the Apache webserver, the MySQL relational database management
system, and the PHP server-side scripting language \citep[see][for
  further details]{Silverman10}.


\section{Analysis}\label{s:analysis}

\subsection{Light Curves}\label{ss:lc}

\begin{figure}
\includegraphics[width=8.5cm]{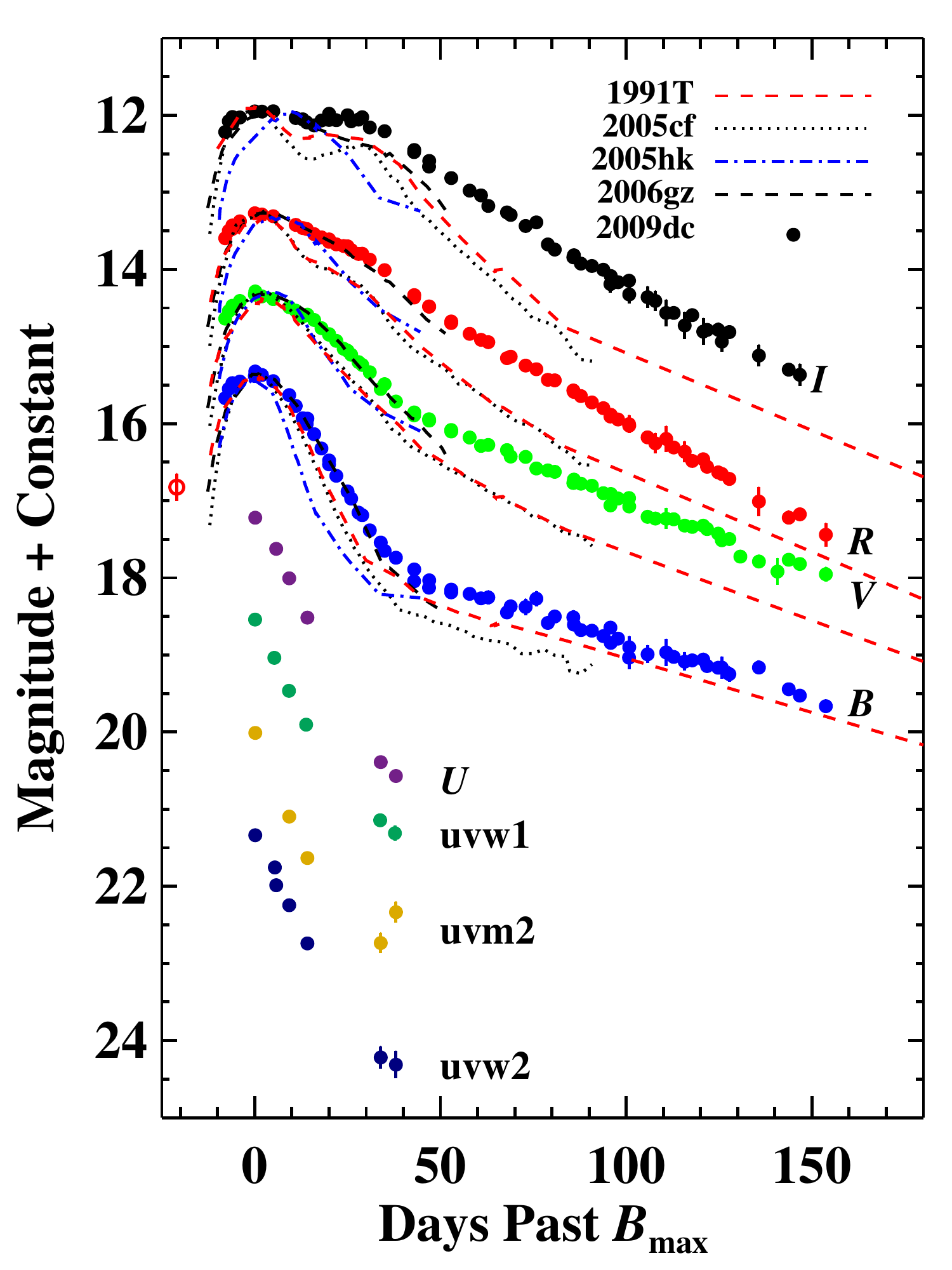}
\caption{Optical and ultraviolet (UV) light curves of SN~2009dc from
  KAIT and the 1-m Lick Nickel telescope. For comparison we also plot
  the optical light curves of the overluminous Type~Ia SN~1991T, the
  normal SN~2005cf, the peculiar SN~2002cx-like SN~2005hk, and another
  SC~SN~Ia candidate SN~2006gz. The light curves of each SN are
  shifted so that $t = 0$~days corresponds to the time of $B_{\rm
    max}$ and reach the same peak magnitude for a given filter.  The
  evolution of \dc\ is atypical compared to other SNe~Ia. \dc\ evolves
  much more slowly and exhibits only a muted secondary maximum in the
  $I$ band.  We include in our $R$-band light curve an early detection
  of \dc\ ({\it red open circle}) in an unfiltered image from the Lick
  Observatory SN Search (LOSS), about 3~weeks before maximum
  brightness.  The data sources for the other SNe are cited in the
  text.}\label{figure:phot}
\end{figure}

We present our final optical \bvri\ light curves of \dc\ in
Figure~\ref{figure:phot}. We include for comparison a ``normal''
Type~Ia SN~2005cf \citep{Wang09:05cf}, another SC~SN~Ia candidate
\gz\ \citep{Hicken07}, the ``standard overluminous'' Type~Ia SN~1991T
\citep{Lira98}, and the peculiar SN~2002cx-like SN~2005hk
\citep{Phillips07}.


The light curves of \dc\ have many features which are noticeably
distinct from most other SNe~Ia. The light curves are much broader
than those of spectroscopically normal SNe~Ia such as SN~2005cf. Even
in comparison to the overluminous SN~1991T, \dc\ evolves significantly
more slowly.The light curve of \gz\ presented by \cite{Hicken07}
provides an good match, although \dc\ appears to have a slower rise
time {\it and} a slower decline.

The absence of a prominent second maximum in the $R$ and $I$ bands is
particularly striking. The secondary maximum in $I$ has been
attributed to the propagation of an ionization front through
iron-group elements (IGEs) as they transition from being doubly ionized
to singly ionized \citep{Kasen06}. Observationally, the strength of
the secondary maximum has been found to correlate with peak
luminosity, powered by Ni decay, with luminous SNe exhibiting more
prominent secondary peaks \citep{Hamuy96,Nobili05}. However, this
feature is weak in \dc, even though it is more luminous than a typical
SN~Ia. \cite{Kasen06} found that significant mixing of $^{56}{\rm Ni}$
in the composition of a SN~Ia can lead to a blending of the two
$I$-band peaks, possibly explaining why a secondary peak is weak in
\dc.

After correcting for Milky Way (MW) extinction \citep[from the dust
  maps of][]{Schlegel98}, we fit a polynomial to the $B$-band light
curve to find that \dc\ peaked on JD $2,454,946.93 \pm 0.2$ (on
2009~Apr.~25.4) at $B_\textrm{max} = 15.05 \pm 0.04$~mag.  These
values are consistent with those derived by \citet{Yamanaka09}.

We also measure $\Delta m_{15} (B)$, the decline in flux from maximum
light to 15~days past maximum in the $B$ band.  Correcting for time
dilation, we find $\Delta m_{15} (B)= 0.72 \pm 0.03$~mag, making
\dc\ one of the most slowly evolving SNe ever discovered and
comparable to other SC~SN~Ia candidates in the
literature. Interestingly, \cite{Yamanaka09} measure $\Delta m_{15}
(B) = 0.65 \pm 0.03$~mag. Comparing the two photometric datasets, we
find discrepancies between measurements of local standard stars of
\about0.15~mag. With over 10 calibrations of the field on photometric
nights, we are confident in our measurements of the field standards
presented in Table~\ref{comp_stars}. While these differences are
troubling, they do not alter the conclusion that \dc\ is a slowly
evolving SN with an extreme value of $\Delta m_{15}(B)$.


The host galaxy of \dc\ is part of the Lick Observatory Supernova
Search \citep[LOSS;][]{li00:kait,Filippenko01}, allowing us to put
strict constraints on the rise time of \dc. Our first detection of
\dc\ is from an unfiltered image on 2009~Apr.~4, when the SN is seen
at $R=18.82 \pm 0.18$~mag, indicating a rise time $>21$~days. This
first detection is included in Figure~\ref{figure:phot} as part of our
$R$-band data \citep[][show that KAIT unfiltered data approximate the
$R$ band]{Li03}. In an image taken on 2009~Mar.~28, there is
no detection of \dc\ down to $R\approx19.3$~mag. With these
constraints we conservatively estimate a rise time of $23 \pm 2$~days.
\cite{Hayden10} recently found an average SN~Ia rise time of $17.38
\pm 0.17$~days using 105 SDSS-II SNe~Ia, with slowly declining SNe
tending to have shorter rise times. The rise time of \dc\ is
significantly longer than the average SN~Ia in their sample, despite
being a slowly declining SN, and therefore does not follow the trend
found for normal SNe~Ia in their sample. \cite{Riess99}, on the other
hand, found an average rise time of $19.5 \pm 0.2$~days for a
``typical'' SN~Ia ($\Delta m_{15}(B) =1.1$~mag), with slowly declining
SNe having slower rise times. Their sample of 10 objects lacks SNe~Ia
with $\Delta m(15)_{B} < 0.95$~mag, making an extrapolation to $\Delta
m_{15}(B) =0.72$~mag to compare with the rise time of \dc\ unreliable.

\begin{figure}
\includegraphics[width=8.5cm]{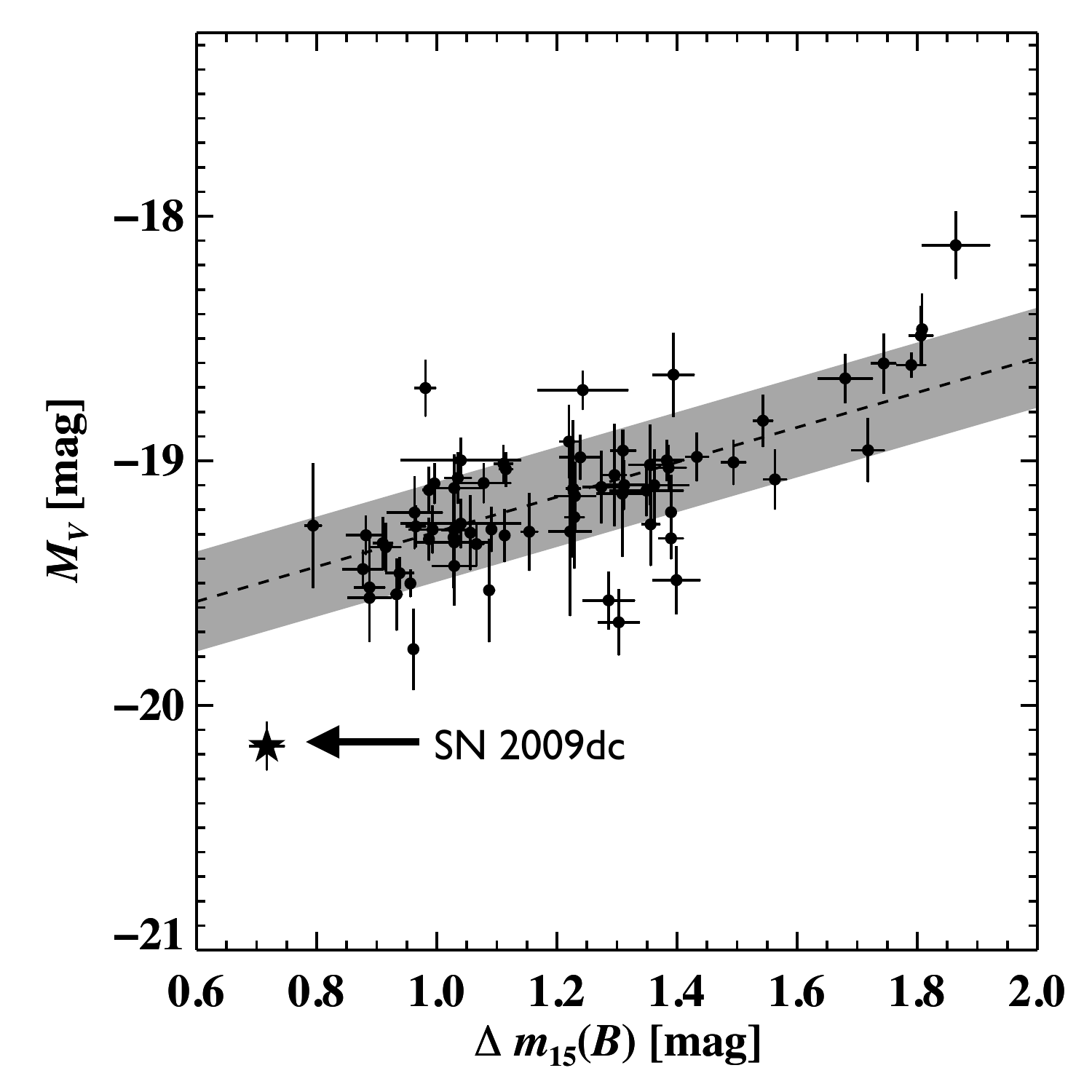}
\caption{$M_{V}$ as a function of $\Delta m_{15}(B)$ for 71 SNe~Ia 
  with $z_{\rm Virgo~infall} > 0.01$ from the LOSS photometry archive 
  \citep[{\it filled circles},][]{Ganeshalingam10:cosmo}, compared
  with \dc\ ({\it filled star}).  Host-galaxy extinction is derived
  using MLCS2k2.v006 \citep{Jha07}, except in the case of \dc\ where we
  adopt $E(B-V)_{\rm host} = 0.1$~mag with $R_{V} = 3.1$ (see
  \S\ref{ss:reddening} for details). The luminosity distance was
  calculated using the $\Lambda$CDM 
  concordant cosmology with $H_{0}=70$~\kms\perMpc, $\Omega_{m} =
  0.27$, and $\Omega_{\Lambda} = 0.73$ \citep{spergel07}. The best-fit
  line to the data, excluding \dc, in the range $0.6 < \Delta
  m_{15}(B) < 1.7$~mag is plotted as a dashed line, with points that
  fall within 1$\sigma$ shaded in grey. \dc\ is clearly an outlier
  that does not follow the luminosity-width
  relation.}  \label{figure:m15b_v_Mv} 
\end{figure}

The cosmological application of SNe~Ia as precise distance indicators
relies on being able to standardize their
luminosity. \cite{phillips93} showed that $\Delta m_{15} (B)$ is well
correlated with luminosity at peak brightness for most SNe~Ia, the
so-called ``Phillips relation''. In Figure~\ref{figure:m15b_v_Mv}, we
plot absolute $V$ magnitude as a function of $\Delta m_{15}(B)$ for 71
SNe~Ia with $z_{\rm Virgo~infall} > 0.01$ from the LOSS photometry
database \citep{Ganeshalingam10:cosmo} and \dc\ to determine whether 
\dc\ follows the Phillips relation. Host-galaxy extinction for the
sample of 71 SNe~Ia is derived using MLCS2k2.v006 with the {\it glosz}
prior, which assumes that the late-time $(B-V)_{t=+35~{\rm d}}$ colour
is indicative of the host-galaxy extinction \citep{Jha07}. For \dc,
the late-time colour evolution differs significantly from that of
normal SNe~Ia; thus, we instead adopt $E(B-V)_{\rm host} =
0.1$~mag, assuming $R_{V} = 3.1$ for the host galaxy
(see \S\ref{ss:reddening} for details). To place the SNe on an
absolute scale, we use the luminosity distance assuming a $\Lambda$CDM
cosmology with $H_{0}=70$~\kms\perMpc. The best-fit line to SNe with
$0.6 < \Delta m_{15}(B) < 1.7$~mag, excluding \dc, is plotted in
Figure~\ref{figure:m15b_v_Mv}, with the area that falls within
1$\sigma$ of the relation shaded in grey. \dc\ is clearly an
overluminous outlier in comparison to SNe~Ia having similar values 
of $\Delta m_{15} (B)$.
Events similar to \dc\ cannot be standardized
using current distance-fitting techniques which rely on
parameterizations of light-curve shape \citep{Howell06}.

We measure the late-time decay of the light curves using a linear
least-squares fit to data in the range \about50--150~days after
maximum. We find a decline of $1.47 \pm 0.04$~mag per 100~days in $B$,
$1.91 \pm 0.02$~mag per 100~days in $V$, $2.78 \pm 0.03$~mag per
100~days in $R$, and $2.87 \pm 0.04$~mag per 100~days in
$I$. \cite{Leibundgut00} find typical decay rates for SNe~Ia of
1.4~mag per 100~days in $B$, 2.8~mag per 100~days in $V$, and 4.2~mag
per 100~days in $I$. \dc\ shows significantly slower decline rates in
$V$ and $I$.

\begin{figure}
\includegraphics[width=8.5cm]{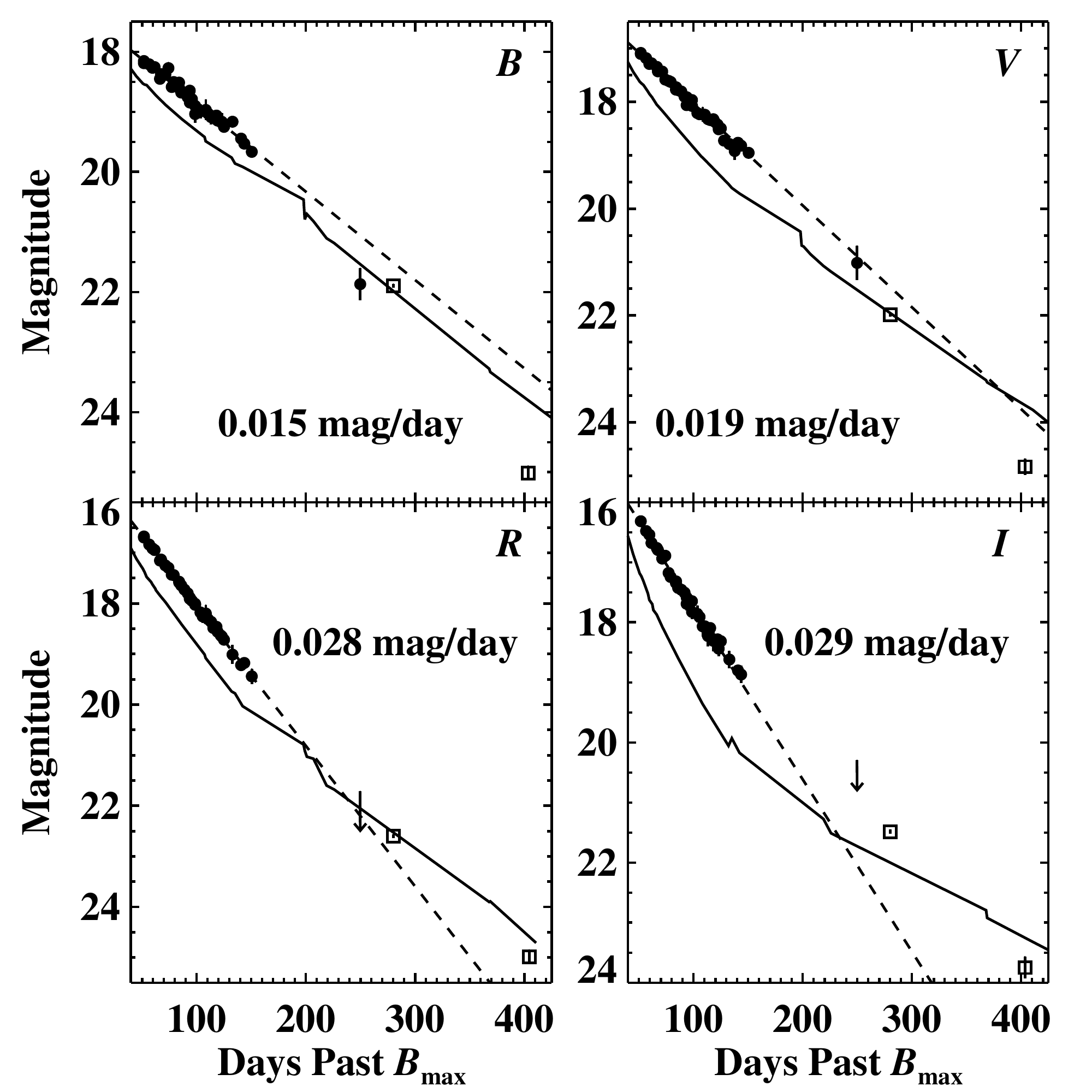}
\caption{The late-time decay of \dc\ from 50 to about 400~days past
  maximum light compared to that of SN~2003du ({\it solid line}), a
  typical SN~Ia.  The light curves are shifted such that $t=0$~days
  corresponds to the time of \bmax, and a constant has been added to
  each SN~2003du light curve to match the peak magnitude of
  \dc. Filled circles are KAIT and Nickel data, whereas open squares are
  LRIS and DEIMOS data. Upper limits are marked with arrows. The LRIS
  $g$-band point is plotted in the $B$-band light curve panel. We 
  fit the decay in \bvri, plotted as a dashed line, using a linear
  least-squares fit to our well-sampled data between 50--150~days past
  maximum, before the SN became projected too close to the
  Sun. Late-time photometry obtained with LRIS at 281~days past
  maximum indicates that the flux in $B$ and $V$ is below what is
  expected from linear extrapolations, but mostly agrees with
  expectations from SN~2003du. However, the flux in $R$ and $I$ is
  larger than what is expected compared to the extrapolations. Our
  $R$-band detection falls on the comparison light curve, while the
  $I$-band detection is brighter than what is expected. \dc\ is
  detected in all bands at \about400~days past maximum, indicating
  that the SN is still active despite the points lying below
  interpolated values from SN~2003du. The data for SN~2003du were
  taken from \citet{Stanishev07}.
}\label{figure:late_time_decay_w_03du} 
\end{figure}

Late-time photometry of \dc\ (Table~\ref{t:late_time_phot}) taken with
the Nickel telescope shows only marginal detections in $B$ and $V$ and
upper limits in $R$ and $I$, while our Keck images, 281~days past
maximum, give clear detections in
\gvri. Figure~\ref{figure:late_time_decay_w_03du} shows the late-time
behaviour of \dc\ in comparison to late-time photometry of SN~2003du,
a typical SN~Ia, taken from \cite{Stanishev07}, and the linear decay
rates found using data \about50--150~days after maximum.  A constant
has been added to each SN~2003du light curve in order to match the
peak magnitude of \dc. Compared to the extrapolations based on the
measured linear decay rates, at 281~days past maximum \dc\ is fainter
by \about\ 0.5~mag in $B$ and $V$, brighter by \about0.5~mag in $R$,
and \about1.5~mag brighter in $I$. In comparison to SN~2003du, \dc\ is
within \about0.15~mag of the interpolated values in $BVR$ and
\about0.5~mag brighter in $I$. We caution that our interpolated values
for SN~2003du suffer from a rather large gap in the light curve
between 225~days and 366~days after maximum. While the late-time
behaviour of \dc\ does not match the linear decay of SN~2003du exactly
in all bands, our detections indicate that it has not undergone an
unexpected drop in luminosity, and it is consistent with the late-time
light curve being powered by $^{56}$Co decay (see \S\ref{ss:nickel}).

The data taken \about400~days past maximum show only marginal
detections in all bands, indicating that the SN is still active despite
the points lying below interpolated values from SN~2003du. The
detections in $B$ and $V$ are clearly below what we would expect from
SN~2003du, while $R$ and $I$ are not too far below expectations. We
address possible reasons for the drop in flux in
\S\ref{sss:late_time}.


\subsection{Colour Evolution}\label{ss:cc}
\begin{figure}
\includegraphics[width=8.5cm]{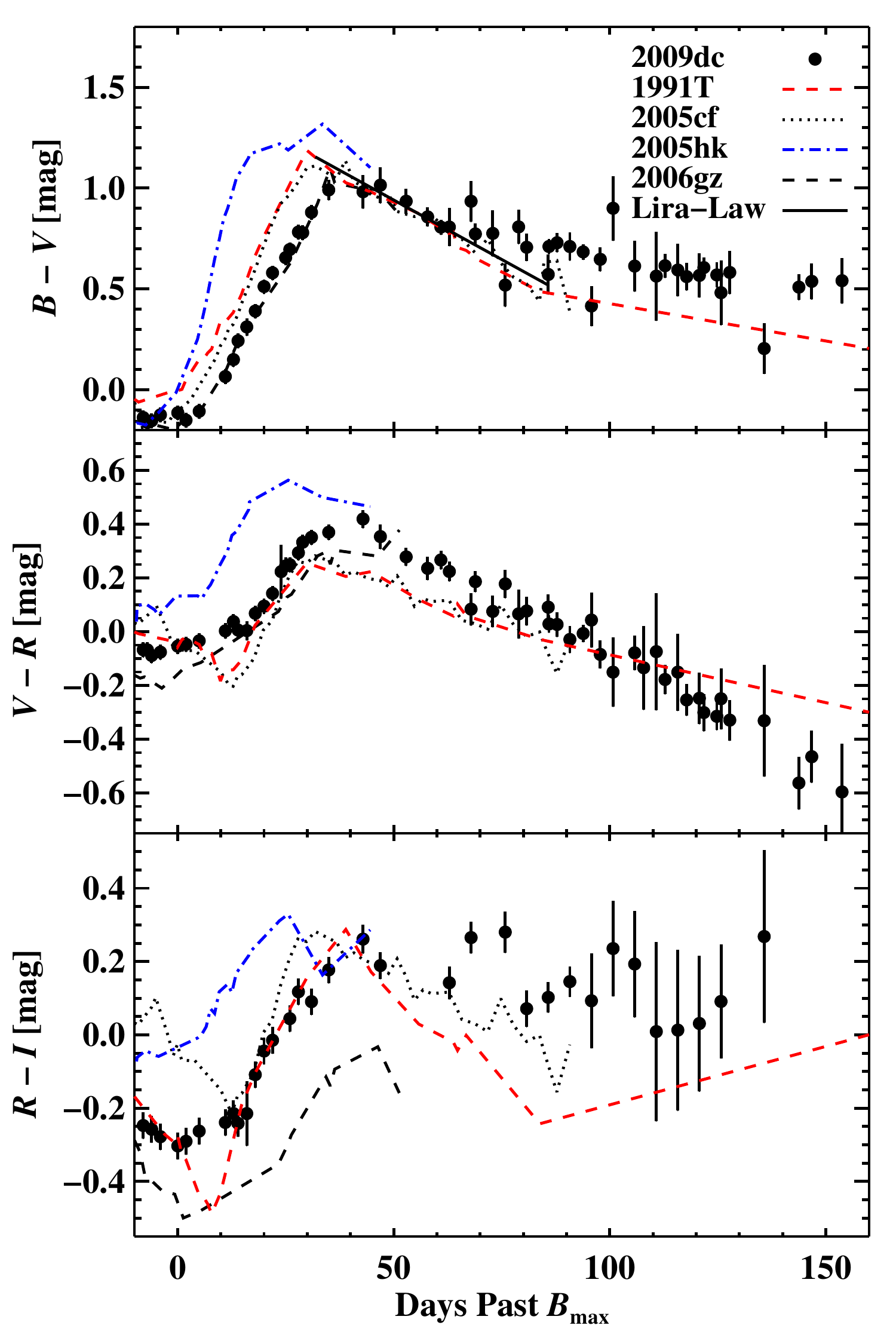}
\caption{\bv\ (top), \vr\ (middle), and \ri\ (bottom) colour curves of
  SN~2009dc. Plotted for comparison are the colour curves of
  SNe~1991T, 2005cf, 2005hk, and 2006gz.  All curves have been
  corrected for MW reddening and host-galaxy extinction. We have not
  included errors in derived host-galaxy extinction for \dc\ which
  will systematically shift curves in one direction.  At early times,
  \dc\ is bluer than SNe~1991T, 2005cf, and 2005hk. The evolution most
  clearly resembles that of SN~2006gz. We plot the Lira law as a solid
  line in the range $35 < t < 85$~days.  \dc\ shows a slower red to
  blue colour evolution compared to the Lira-Phillips relation,
  especially after $t=70$~days. The data sources are cited in the
  text.} \label{figure:color}
\end{figure}

The colour evolution in \bv, \vr, and \ri\ for \dc\ in comparison to
SNe~2006gz, 2005cf, 1991T, and 2005hk is displayed in
Figure~\ref{figure:color}. All SNe have been corrected for MW
extinction using reddening derived from the dust maps of
\citet{Schlegel98}. We have also corrected for host-galaxy extinction,
assuming $R_{V} = 3.1$, using the following reported values of
$E(B-V)_{\rm{host}}$: 0.13~mag for SN~1991T \citep{Lira98}, 0.10~mag
for SN~2005cf \citep{Wang09:05cf}, 0.09~mag for SN~2005hk
\citep{Phillips07}, 0.18~mag for SN~2006gz \citep{Hicken07}, and
0.10~mag for \dc.

The colour curves indicate that \dc\ was a particularly blue SN Ia
even compared to the prototypical overluminous SN~1991T. In \bv,
\dc\ remains bluer than SNe~1991T, 2005cf, and 2005hk until
\about50~days past maximum light. The colour curves for \dc\ are most
similar to those of \gz, another SC~SN~Ia candidate. At $t>50$~days,
the \bv\ colour of \dc\ becomes redder than that of SN~1991T and
SN~2005cf.

\cite{lira96} showed that SNe with low host-galaxy extinction had
similar \bv\ colour evolution between $t$ = +30 to +90~days
independent of light-curve shape (the ``Lira-Phillips
relation''). \cite{Hicken07} used this relationship to derive the
host-galaxy extinction for \gz. However, a comparison between the
slope of the Lira-Phillips relation and the \bv\ colour evolution of
\dc\ shows disagreement. Adopting the relation derived using 6
low-reddening SNe~Ia by \cite{Phillips99} and fitting for a constant
$E(\bv)_\textrm{host}$ between $t$ = +35 and +90~days, we find
$E(\bv)_\textrm{host} = 0.26 \pm 0.04$~mag, with $\chi^2=41$ for 15
degrees of freedom. If we restrict our fit to between $t$ = +35 and
+70~days, we find a better fit with $E(\bv)_\textrm{host} = 0.19 \pm
0.04$~mag; $\chi^2=5$ for 7 degrees of freedom. We include the fit to
the Lira-Phillips relation in Figure~\ref{figure:color}.

\subsection{Pre-Maximum Spectrum}\label{ss:premaxspec}

On 2009~Apr.~16.22, \about9~days before $B$-band maximum,
\citet{Harutyunyan09} obtained a spectrum of \dc\ which showed
``prominent \ion{C}{II} lines and a blue continuum'', and they noted
that the spectrum resembled pre-maximum spectra of
\gz\ \citep{Hicken07}, but with a much lower expansion velocity.  We
obtained a spectrum \about2~days later which confirmed the spectral
peculiarities noted by \citet{Harutyunyan09}.

\begin{figure*}
\includegraphics[width=18.5cm]{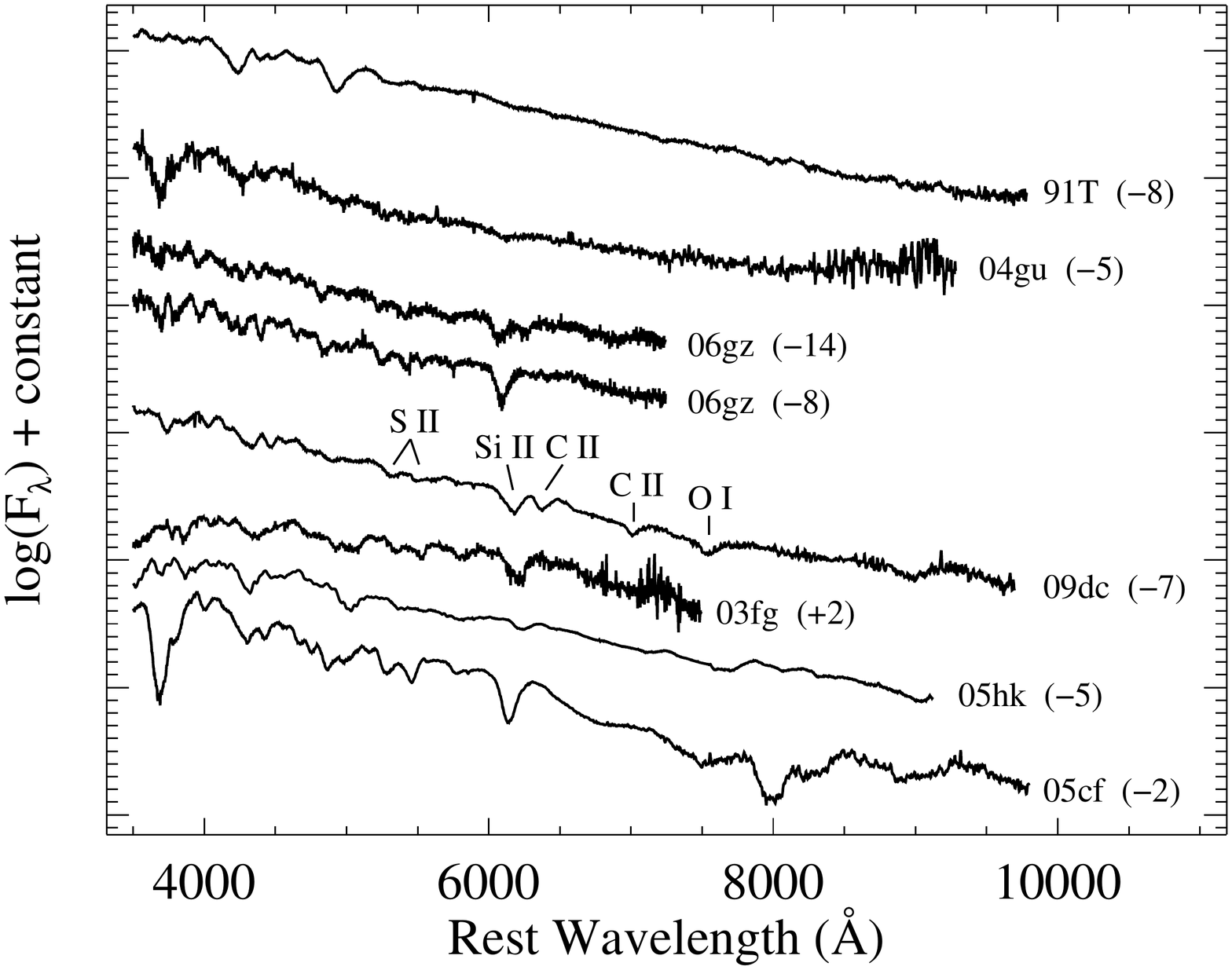}
\caption{Our pre-maximum spectrum of \dc\ (and various comparison SNe)
  with a few major lines identified and days relative to maximum light
  indicated for each spectrum (in parentheses).  From top to bottom,
  the SNe are the ``standard overluminous'' Type~Ia SN~1991T
  \citep{Filippenko92}, a possible SC~SN~Ia SN~2004gu \citep[][though
    this spectrum is from our own database]{Contreras09}, another
  possible SC~SN~Ia \gz\ \citep{Hicken07} at two different epochs,
  \dc\ (this work), yet another possible SC~SN~Ia
  \fg\ \citep{Howell06}, the SN~2002cx-like peculiar SN~2005hk
  \citep{Chornock06}, and finally the ``standard normal'' Type~Ia
  SN~2005cf \citep{Wang09:05cf}. All spectra have had their
  host-galaxy recession velocities removed and have been dereddened
  according to the values presented in their respective references.
  Note that \ion{Si}{II}, \ion{S}{II}, and \ion{O}{I} are often found
  in near-maximum spectra of SNe~Ia, but \ion{C}{II} is
  not.}\label{f:early_spec}
\end{figure*}

Our pre-maximum spectrum is shown in Figure~\ref{f:early_spec}, where
we compare \dc\ to other possible SC~SNe~Ia: \gz\ \citep{Hicken07}, 
\fg\ \citep{Howell06}, and 
SN~2004gu \citep[][though the displayed spectrum is from our own
database]{Contreras09}. 
In addition, we show for comparison spectra from similar epochs of the
``standard overluminous'' Type~Ia SN~1991T \citep{Filippenko92}, the
SN~2002cx-like peculiar SN~2005hk \citep{Chornock06}, and the
``standard normal'' Type~Ia SN~2005cf \citep{Wang09:05cf}.

\dc\ has strong \ion{Si}{II} and \ion{S}{II} lines, and appears to
also contain \ion{O}{I}, all of which are usually found in
near-maximum spectra of SNe~Ia. However, there are also two apparent
\ion{C}{II} lines (labeled in Fig.~\ref{f:early_spec} and discussed in
\S\ref{sss:carbon}) seen in \dc.  The existence (and strength) of
these features is one reason that \dc\ was immediately classified as a
possible SC~SN~Ia \citep{Harutyunyan09}, and one way used to
distinguish it from other subtypes of SNe~Ia.

Spectroscopically, \dc\ does not seem to resemble any of SNe~1991T,
2005hk, or 2005cf.  Although \gz\ has weak \ion{C}{II} features at
14~days before maximum, by 8~days before maximum they have almost
completely disappeared (whereas \dc\ has strong \ion{C}{II} at 7~days
before maximum).  In addition, it is apparent from
Figure~\ref{f:early_spec} that the expansion velocity of \gz\ is much
larger than that of \dc\ (based on the \ion{Si}{II} $\lambda$6355
feature, for example; see \S\ref{sss:silicon} for further details).
Finally, \fg\ does seem to resemble \dc, even though the spectrum of
\fg\ has low signal-to-noise ratio (S/N) and the \ion{C}{II} features
are not nearly as prominent in \fg\ as they are in \dc.  It is
possible that this is due to the fact that the spectrum of \fg\ was
taken 9~days later (relative to maximum light) than our spectrum of
\dc.

\subsubsection{\ion{Si}{II}}\label{sss:silicon}

The tell-tale spectroscopic signature of a SN~Ia is broad absorption
from \ion{Si}{II} $\lambda$6355 \citep[e.g.,][]{Filippenko97}, and
\dc\ is no different. One thing that {\it is} interesting about the
\ion{Si}{II} feature in \dc\ is its extremely low velocity. The
minimum of the \ion{Si}{II} $\lambda$6355 absorption feature was found
to be blueshifted by $8700$~\kms\ about 9~days before maximum brightness
\citep{Harutyunyan09}, and we measure the feature to be blueshifted by
\about8600~\kms\ in our spectrum taken 7~days before maximum.
\citet{Yamanaka09} report it to be blueshifted by $8000$~\kms\ in
their spectrum 3~days before maximum, decreasing to $6000$~\kms\ by
25~days after maximum.

These velocities are much lower than the average photospheric velocity
near maximum for SNe~Ia of \about11,000~\kms\ \citep{Wang09}.  In
fact, according to \citet{Wang09}, the \ion{Si}{II} velocities seen in
\dc\ are approximately 8$\sigma$ below the average.\footnote{Note that
  they consider SNe~Ia with photospheric velocities greater than
  3$\sigma$ above the average to be ``high-velocity SNe~Ia.''}

Interestingly, these velocities {\it are} similar to those of two
other possible SC~SNe~Ia, \fg\ and
\snif\ \citep[][respectively]{Howell06,Scalzo10}.
However, two other objects that were claimed to be possible SC~SNe~Ia
show much more normal \ion{Si}{II} velocities. \citet{Hicken07} report
expansion velocities near maximum light of about
11,000--13,000~\kms\ for \gz.
SN~2004gu, which was compared to \gz\ by \citet{Contreras09}, also
shows a normal photospheric velocity of \about11,500~\kms\ at 5~days
before maximum (this work) slowing to \about11,000~\kms\ at 3~days
past maximum.\footnote{This second velocity has a large uncertainty
  since it was measured by approximating the minimum of the
  \ion{Si}{II} absorption feature by eye from Figure~3 of
  \citet{Contreras09}.}

Figure~4 of \citet{Yamanaka09} and Figure~4 of \citet{Scalzo10} both
present the velocity evolution of the \ion{Si}{II} $\lambda$6355
feature for various possible SC~SNe~Ia and other comparison SNe.  Our
new early-time data point (i.e., \about8600~\kms\ at $t\approx
-7$~days) is quite consistent with the rest of the published
\dc\ data.  Moreover, we detect \ion{Si}{II} $\lambda$6355 in
our next spectrum taken 35~days past maximum at a blueshifted velocity
of about 5500~\kms\ (see \S\ref{sss:onemonth} for further details).
This yields a velocity gradient of \about74~\kms~day$^{-1}$, which is
over twice as large as that found for
\snif\ \citep[34~\kms~day$^{-1}$;][]{Scalzo10}.  Furthermore, combining
our measurements with all previously published values of the expansion
velocity of \dc\ \citep{Harutyunyan09,Yamanaka09,Tanaka09}, we see no
strong evidence for a velocity ``plateau'' near maximum light like the
one seen in \snif\ \citep{Scalzo10}.


We measure the equivalent width (EW) of the \ion{Si}{II} $\lambda$6355
feature to be \about40~\AA\ in our spectrum of \dc\ obtained 7~days
before maximum.  This is similar to the EW of the same line in \gz\ at
similar epochs \citep{Hicken07}, and both are well below the average
\ion{Si}{II} EW for normal SNe~Ia \citep[though they are comparable to
  the overluminous SN~1991T-like SNe~Ia, e.g.,][]{Hachinger06,Wang09}.

\subsubsection{\ion{C}{II}}\label{sss:carbon}

As ubiquitous as \ion{Si}{II} is in the spectra of SNe~Ia, \ion{C}{II}
is nearly as rare. In a few cases, mainly at very early times, weak
\ion{C}{II} has been detected in optical spectra of SNe~Ia
\citep[e.g.,][]{Branch03,Thomas07,Tanaka08,Foley10}. Candidate 
SC~SNe~Ia, on the other hand, exhibit strong \ion{C}{II} in their 
near-maximum spectra \citep[][]{Howell06,Hicken07,Scalzo10}.

In our spectrum of \dc\ obtained 7~days before maximum brightness, we
detect absorption from \ion{C}{II} $\lambda$6580 and $\lambda$7234
\citep[both of which were also detected in a spectrum obtained the same
  day by][]{Marion09}. The minimum of the absorption from \ion{C}{II}
$\lambda$6580 is blueshifted by about 9500~\kms\ in our pre-maximum
spectrum. The expansion velocity of this feature is
\about8500~\kms\ at 3~days before maximum, slowing to about
7000~\kms\ at 3~days after maximum, and \about6700~\kms\ by 6~days
after maximum \citep{Yamanaka09,Tanaka09}. By about 18 days past
maximum this feature seems to have completely disappeared, though it
is difficult to be sure given the low S/N spectrum seen in Figure~3 of
\citet[][]{Yamanaka09}.  However, it is clear from the same figure
that by 25~days after maximum, the feature is most definitely gone. It
is also undetected in our spectrum obtained 35~days past maximum (see
\S\ref{ss:postmaxspec}).

Also, from our pre-maximum spectrum of \dc, we measure the minimum of
the \ion{C}{II} $\lambda$7234 absorption to be \about9400~\kms, which
is in good agreement with the velocity of the \ion{C}{II}
$\lambda$6580 feature that we calculate above.  This feature is still
apparent at about 6~days past maximum, near a velocity of
\about6600~\kms\ \citep[see Fig.~1 of][though the feature is not
  marked]{Tanaka09}, but it has also likely disappeared by 18~days
past maximum \citep[][Fig.~3]{Yamanaka09} and is certainly gone by
35~days past maximum (see \S\ref{ss:postmaxspec}).

For \ion{C}{II}, the velocities seen in \dc\ once again match those of
\fg\ at \about2~days past maximum \citep[][Fig.~3 and Supplementary
  Information]{Howell06}, though only the \ion{C}{II} $\lambda$4267
line (at an expansion velocity of \about8300~\kms) is clearly
detected. On the other hand, \snif\ appears to have both the \ion{C}{II}
$\lambda$6580 and $\lambda$7234 features in its spectrum from 5~days
past maximum \citep{Scalzo10}.

A spectrum of \gz\ at 13~days before maximum shows both the
\ion{C}{II} $\lambda$6580 and \ion{C}{II} $\lambda$7234 lines with
expansion velocities near 15,500~\kms. However, both features are
effectively undetectable 4~days later \citep{Hicken07}.  We also note
that neither line is seen in spectra of SN~2004gu at 5~days before
maximum (see Fig.~\ref{f:early_spec}, although this spectrum has low
S/N) or at 3~days after maximum \citep{Contreras09}.\footnote{The
  \ion{C}{II} $\lambda$7234 line, if present, would be at the red end
  of this spectrum.}

For \dc, \citet{Marion09} reported no detectable absorption from
\ion{C}{II} $\lambda$4745 in their spectrum obtained 7~days before
maximum.  However, in our spectrum taken on the same day, we {\it do}
see evidence for absorption from this transition, as well as
absorption from \ion{C}{II} $\lambda$4267 (see \S\ref{sss:synow}).
Note that both of these features are slightly blended with absorption
from other species, which is what most likely led \citet{Marion09} to
conclude that the features were not present.


The EW of the \ion{C}{II} $\lambda$6580 and $\lambda$7234 absorptions,
as calculated from our spectrum taken 7~days before maximum, are
\about19~\AA\ and \about13~\AA, respectively.  \citet{Hicken07} report
that the \ion{C}{II} $\lambda$6580 feature in \gz\ 14~days before
maximum had an EW that was slightly higher than these values (25 \AA).
Both of these SNe have much stronger \ion{C}{II} lines than typical
SNe~Ia.  For example, one of the strongest \ion{C}{II} features ever
seen in a normal SN~Ia was in SN~2006D, where \ion{C}{II}
$\lambda$6580 was detected with an EW of merely
7~\AA\ \citep{Thomas07}.

\subsubsection{\ion{Ca}{II}}\label{sss:calcium}

In our pre-maximum spectrum (as well as most of our post-maximum
spectra; see \S\ref{ss:postmaxspec}) we detect narrow, as well as
broad, absorption from \ion{Ca}{II}~H\&K. The narrow components are at
the redshift of \dc\ and its host galaxy.  At all epochs these lines
are just at our instrumental resolution limit; thus, calculating an
accurate FWHM is nearly impossible.  In addition, the H\&K lines occur
in a part of the spectrum where numerous broad absorption features
blend together, so defining a local continuum is difficult, 
exacerbating the problem of measuring reliable line strengths.
Despite all this, we attempt to measure the strength of this narrow
feature.  We calculate that the \ion{Ca}{II}~H\&K lines appear to have
FWHM ranging from about 100--300~\kms.  
However, the line strengths are consistent with no change during these
epochs, given our rather large uncertainties.

Some SNe~Ia exhibit narrow, time-variable \ion{Na}{I}~D absorption
lines with unchanging, narrow \ion{Ca}{II}~H\&K absorption
\citep[e.g.,][]{Simon09}. However, time-variable \ion{Ca}{II}~H\&K
absorption has not been seen.  \dc\ does show weak, narrow absorption
from \ion{Na}{I}~D, but it is far too weak to determine if variability
in the line strengths exists (see \S\ref{ss:reddening}).  We cannot
claim to have detected narrow, time-variable \ion{Ca}{II}~H\&K
absorption in \dc, but such variability should be searched for in
other, future SC~SNe~Ia.

The narrow \ion{Ca}{II}~H\&K may arise from calcium-rich interstellar
material (ISM) in the host galaxy of \dc, UGC~10064, and in fact this
feature is strong in a spectrum of the core of the galaxy
\citep{Abazajian09}.  However, since \dc\ is relatively far from the
host's nucleus, this seems somewhat unlikely.  The alternative is that
the calcium-rich material which gives rise to the strong
\ion{Ca}{II}~H\&K line is in close proximity to the SN site.  This
means that \dc\ likely exploded in a region of calcium-rich ISM, or
that the progenitor star of \dc\ itself had calcium-rich circumstellar
material.  However, which of these two possibilities best reflects the
true situation and how this may affect the SN itself are beyond the
scope of this paper.

Another \ion{Ca}{II} feature commonly seen in SNe~Ia is the broad
\ion{Ca}{II} near-IR triplet. However, we see no hint of this line
in our spectrum obtained 7~days before maximum.  On the other hand,
\citet{Tanaka09} clearly detect this feature near 8400~\AA\ in their
spectropolarimetric observations taken \about6~days after maximum.
Not only is the line apparent in their total-flux spectrum, but it
is the most polarized spectral feature they detect, with a total
polarization of $\sim 0.7$\%.  This, along with the high polarization
of \ion{Si}{II} $\lambda$6355 and the relatively low continuum
polarization, led \citet{Tanaka09} to conclude that \dc\ had a roughly
spherical photosphere \citep[as is the case for most normal SNe~Ia,
  e.g.,][]{Leonard05,Chornock08} with a somewhat clumpy distribution
of intermediate-mass elements (IMEs).

\subsubsection{Other Species and the SYNOW Fit}\label{sss:synow}

To determine which other species are present in our pre-maximum
spectrum of \dc, we use the spectrum-synthesis code SYNOW
\citep{Synow}.  SYNOW is a parameterized resonance-scattering code
which allows for the adjustment of optical depths, temperatures, and
velocities in order to help identify spectral features seen in SNe.

Before fitting, we deredden our pre-maximum spectrum of \dc\ using
$E(\bv)_\textrm{MW}=0.070$~mag, $E(\bv)_\textrm{host}=0.1$~mag (see
\S\ref{ss:reddening} for more information on how we derive these
values), $R_V = 3.1$, and the reddening curve of \citet{Cardelli89}.
We also remove the host-galaxy recession velocity ($cz = 6300 \pm
300$~\kms, as determined from the narrow \ion{Ca}{II} H\&K absorption
in our spectrum) before comparing with the output of SYNOW.
 
\begin{figure*}
\includegraphics[width=18.5cm]{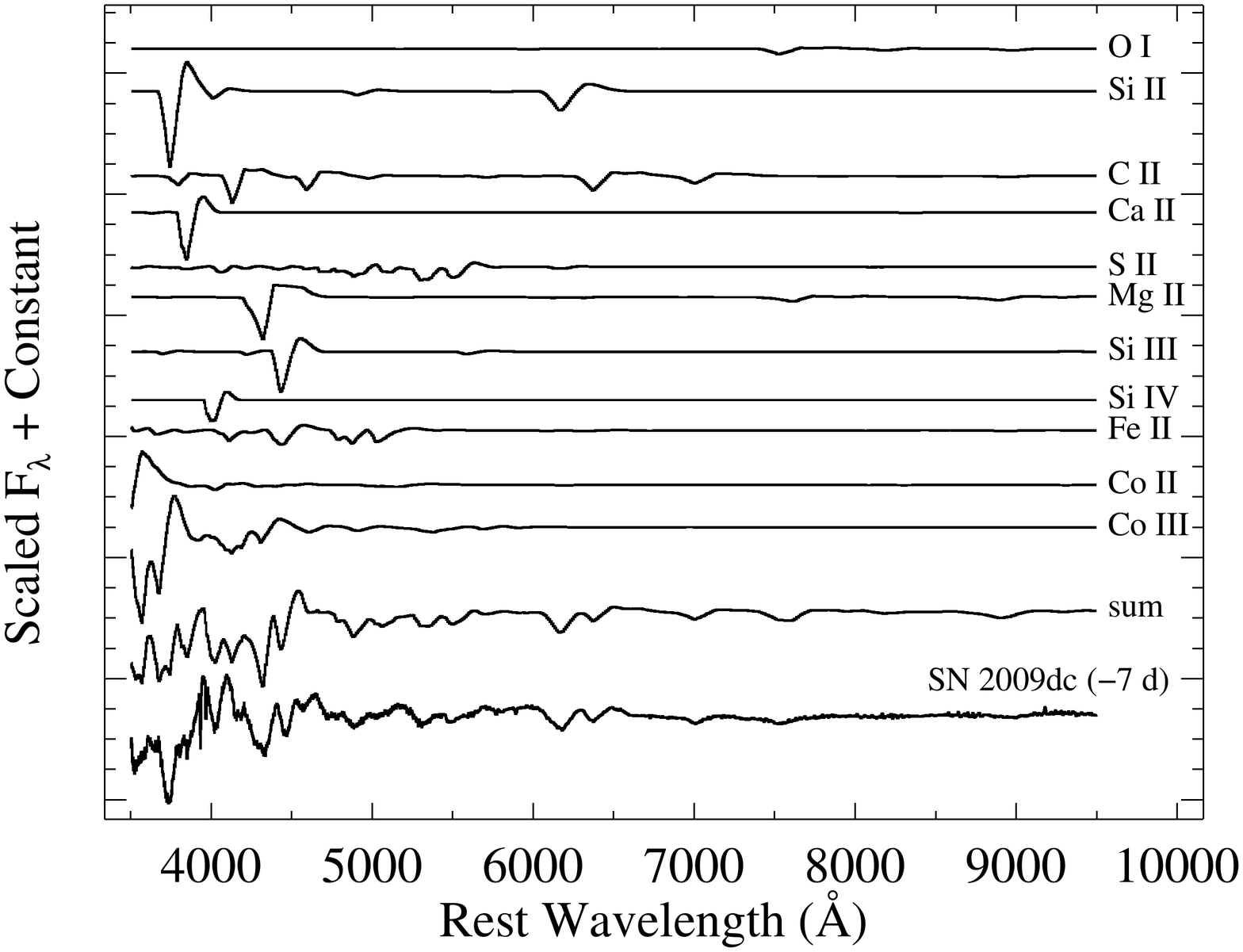}
\caption{The top eleven spectra show the constituent species in our
  SYNOW fit.  The second from bottom spectrum is the sum of the top
  eleven spectra.  The bottom spectrum is our pre-maximum spectrum of
  \dc\ (dereddened by $E(\bv)_\textrm{MW}=0.070$~mag and
  $E(\bv)_\textrm{host}=0.1$~mag with $R_V = 3.1$). We also remove the
  host-galaxy recession velocity ($cz = 6300 \pm 300$~\kms).  Finally,
  we remove the underlying 20,000~K blackbody continuum from all
  spectra shown.}\label{f:synow}
\end{figure*}

Our derived SYNOW fit is compared to our actual pre-maximum spectrum
in Figure~\ref{f:synow}. Also shown are the spectral features from
each of the individual species that were used in the final fit. Our
SYNOW fit faithfully reproduces the vast majority of the features seen
in our pre-maximum spectrum of \dc, but a few features are not matched
exactly (specifically, the ones near 3700~\AA, 4100~\AA, and
4600~\AA).

In addition to the species mentioned above (\ion{Si}{II}, \ion{C}{II},
and weak \ion{Ca}{II}), we detect \ion{O}{I}, \ion{S}{II},
\ion{Mg}{II}, \ion{Si}{III}, \ion{Si}{IV}, \ion{Fe}{II}, \ion{Co}{II},
and \ion{Co}{III}. All of these ions (except \ion{C}{II}, as mentioned
previously) are often found in early-time spectra of SNe~Ia, though
\ion{O}{I} may be weak in some overluminous SNe~Ia
\citep[e.g.,][]{Filippenko97}. The detection of \ion{C}{II} in the
spectra of \dc\ (see \S\ref{sss:carbon}) implies that it had a
significant amount of unburned material in its ejecta, and thus the
oxygen seen in \dc\ is most likely from this unburned material as
well.

Our SYNOW fit has a photospheric velocity of 9000~\kms, with maximum
velocities of 10,000--15,000~\kms\ depending on the ion.  This value
matches our derived velocities from the \ion{Si}{II} and \ion{C}{II}
absorption features (see \S\ref{sss:silicon} and \S\ref{sss:carbon},
respectively).  Since the early-time spectra of \dc\ and \fg\ greatly
resemble each other, it is not surprising that these parameters are
similar to those found in a SYNOW fit of \fg\ \citep[][Supplementary
  Information]{Howell06}. Our fit requires a slightly higher
photospheric velocity (they use 8000~\kms), but this seems reasonable
since our spectrum was obtained 7~days {\it before} maximum and the
spectrum of \fg\ fit by \citet{Howell06} is from \about2~days {\it
  after} maximum.

All ions in our SYNOW fit had an excitation temperature around 10,000
K except \ion{C}{II}, which was raised to 20,000~K in order to get the
relative line strengths to match the data. This is lower than what was
calculated by \citet{Howell06} for \fg.  They required an excitation
temperature for \ion{C}{II} of 35,000~K to match their spectrum, while
we obtain a temperature that is less than 60\% of that value.

Finally, we require an underlying blackbody temperature of 20,000~K in
order to reproduce the overall continuum shape of \dc.  This is hot
for a SN~Ia, even at early times, and is indicative of the production
of a large amount of \nic\ \citep{Nugent95}.  Blackbody temperatures
in the range of 10,000--15,000~K seem to be more common for normal
SNe~Ia \citep[e.g.,][]{Patat96}, with the overluminous Type~Ia
SN~1991T having temperatures at the high end of that range
\citep{Mazzali95}.  \citet{Howell06} used a blackbody temperature of
9000~K to fit \fg, which does not seem abnormally high for a SN~Ia
2~days past maximum.  Even though this value is a bit lower than our
derived blackbody temperature, it is not too surprising since our
spectrum of \dc\ was obtained about 9~days earlier (relative to
maximum light) than their spectrum of \fg.

\subsection{Post-Maximum Spectra}\label{ss:postmaxspec}

\begin{figure*}
\includegraphics[width=18.5cm]{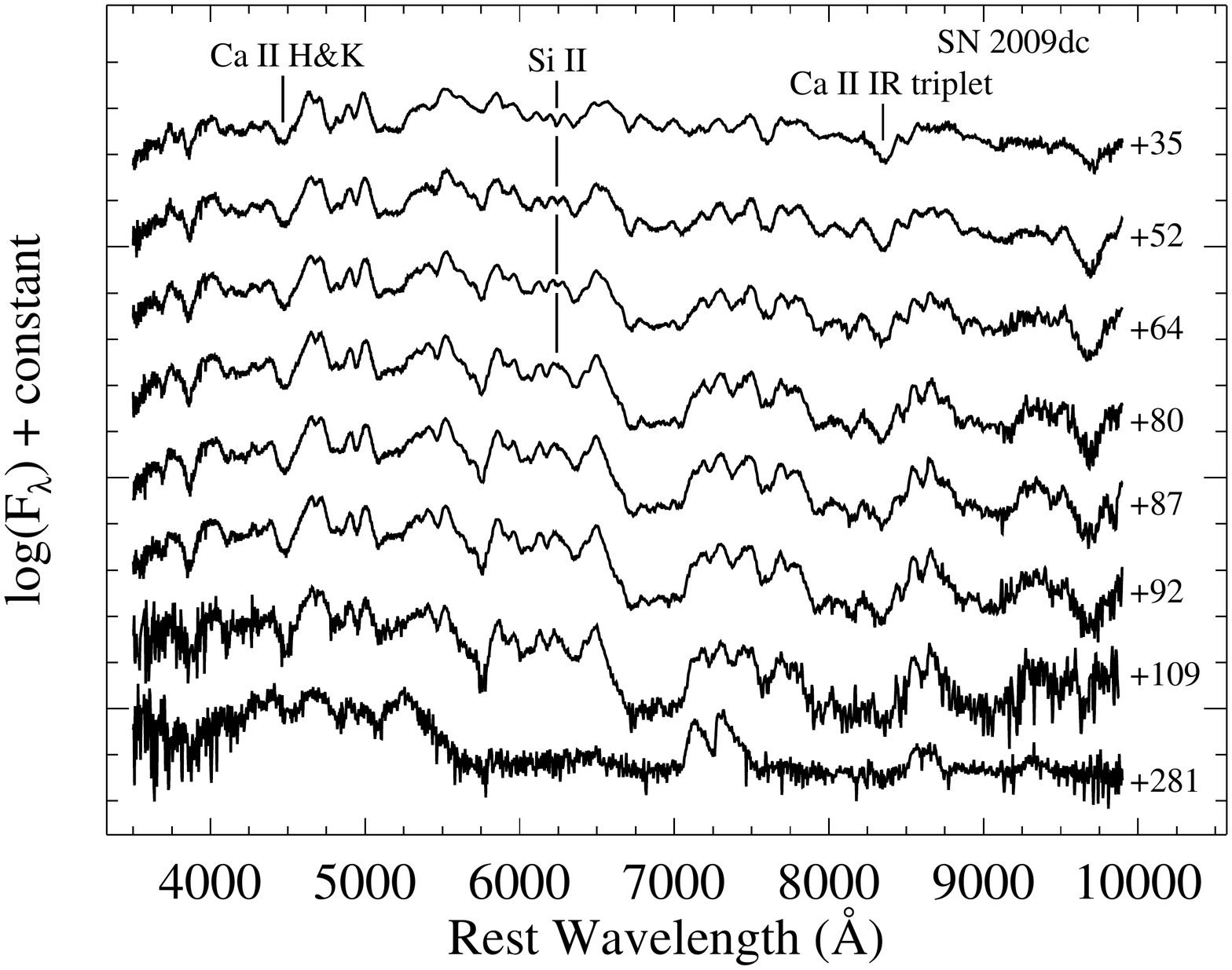}
\caption{Our post-maximum spectra of \dc, with days relative to
  maximum light indicated for each spectrum. Important features
  discussed in the text are labeled. All spectra have been dereddened
  by $E(\bv)_\textrm{MW}=0.070$~mag and $E(\bv)_\textrm{host}=0.1$~mag
  with $R_V = 3.1$. We have also removed the host-galaxy recession
  velocity ($cz = 6300 \pm 300$~\kms).}\label{f:post-max}
\end{figure*}

The rest of our spectral data of \dc\ (i.e., all of our post-maximum
spectra) are presented in Figure~\ref{f:post-max}. Note that
during our observation on 2009~Aug.~14.2 (day 109), there was smoke and
ash in the sky throughout the night (from a fire near the
observatory), and thus the continuum shape is less accurate (due to
the nonstandard extinction caused by the smoke) than in the other
observations.  

\subsubsection{1~Month After Maximum}\label{sss:onemonth}

\begin{figure*}
\includegraphics[width=18.5cm]{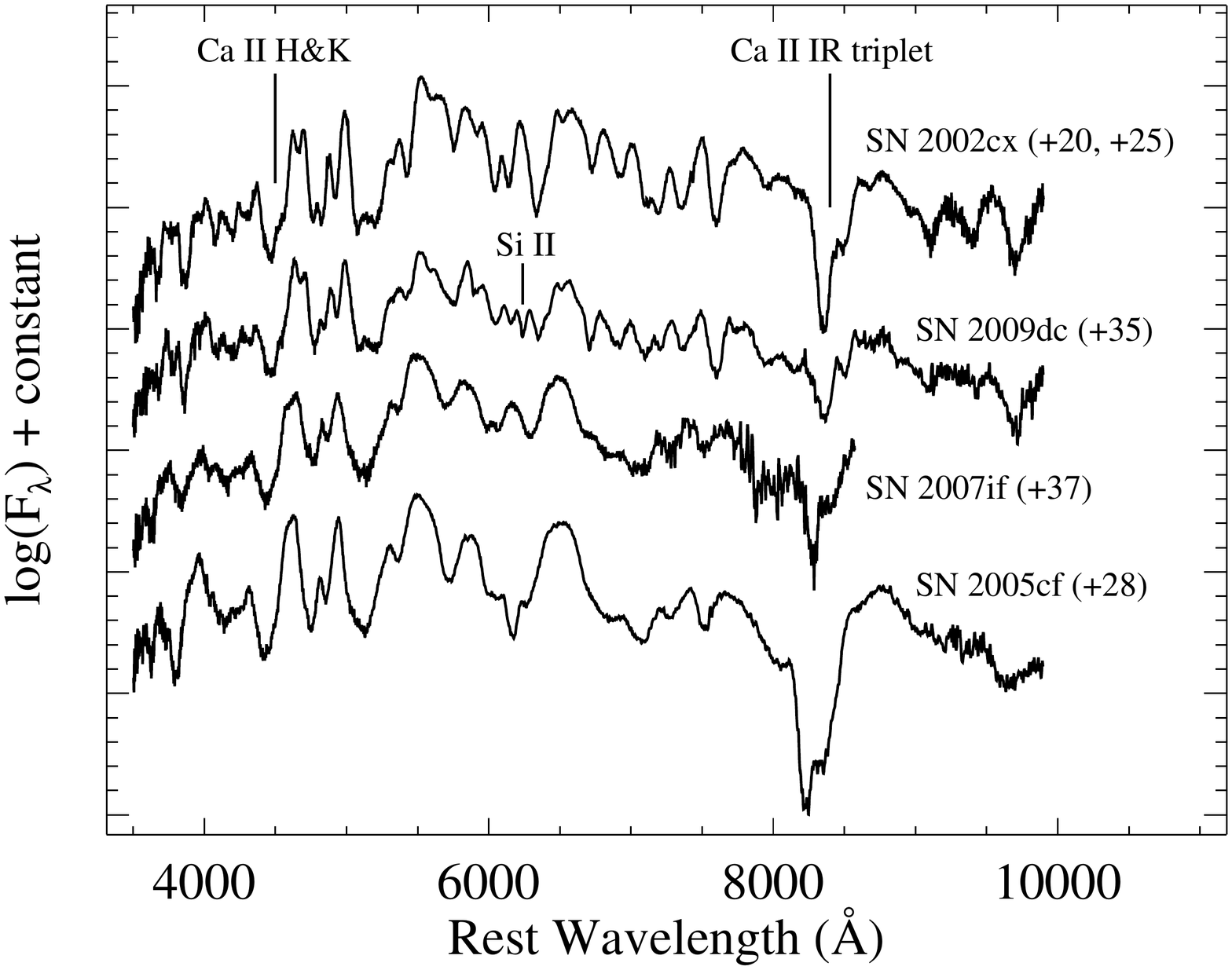}
\caption{Our spectrum of \dc\ 35~days past maximum and a few
  comparison SNe, with days relative to maximum light indicated for
  each spectrum (in parentheses).  From top to bottom, the SNe are the
  peculiar SN~2002cx \citep[][where we have combined their spectra
    from 20 and 25~days past maximum]{Li03:02cx}, \dc\ (this work),
  another possible SC~SN~Ia \snif\ \citep[][though the spectrum shown
    is from our own database]{Scalzo10}, and the ``standard normal''
  Type~Ia SN~2005cf \citep{Wang09:05cf}. Important features discussed
  in the text are labeled. All spectra have had their host-galaxy
  recession velocities removed and have been dereddened according to
  the values presented in their respective
  references.}\label{f:one_month_comparison}
\end{figure*}

Our day 35 (2009~May~31.3) spectrum of \dc\ is shown in
Figure~\ref{f:one_month_comparison}, along with spectra of the
peculiar SN~2002cx \citep{Li03:02cx}, another possible SC~SN~Ia
\snif\ \citep[][though the displayed spectrum is from our own
  database]{Scalzo10}, and the ``standard normal'' Type~Ia SN~2005cf
\citep{Wang09:05cf}.  While there are some spectroscopic similarities
between \dc\ and SN~2005cf at this epoch, there are obvious
differences as well.  Most notably, there are many more medium-width
features in the spectrum of \dc\ compared with SN~2005cf, and the
features that do clearly match are at much lower velocities in \dc.
This is the same as what was seen in the pre-maximum spectra
(\S\ref{ss:premaxspec}).  Furthermore, the \ion{Ca}{II}~H\&K feature
as well as the \ion{Ca}{II} near-IR triplet are much stronger in
SN~2005cf than they are in \dc, again consistent with the pre-maximum
spectrum of \dc\ (\S\ref{sss:calcium}).

The possible SC~SN~Ia \snif\ appears to be a reasonable match to the
spectrum of \dc, but there are some significant differences.
Even though weak \ion{Ca}{II} features are detected in both of these
objects, there are still a few medium-width features in \dc\
that are not seen in \snif.  In addition, the velocities of some (but
not all) of the features in \snif\ are larger than those in
\dc\ by \about1000~\kms.  However, \snif\ appears to match
SN~2005cf well at this epoch, even though its expansion velocity
is still not as large as that of SN~2005cf, and \snif\ is missing
the \ion{Si}{II} $\lambda$6355 absorption seen in SN~2005cf.

The best match to \dc\ at this epoch appears to be the peculiar
SN~2002cx at an age about 5--10~days younger than \dc. A detailed
discussion of the comparison between peculiar SN~2002cx-like SNe and
\dc\ can be found in \S\ref{sss:cx}.  Since nearly every single
spectral feature of \dc\ at this epoch matches those in SN~2002cx, we
use the line identifications of \citet{Li03:02cx} \citep[which mainly came
  originally from][]{Mazzali97} for \dc.  The spectra at this time are
dominated by various multiplets of \ion{Fe}{II}, although \ion{Ca}{II}
(as mentioned above), \ion{Co}{II}, and \ion{Na}{I} are also present.
There is evidence for \ion{O}{I} $\lambda$7773 and \ion{O}{I}
$\lambda$9264 (which may give rise to the features seen near
7600~\AA\ and 9100~\AA, respectively), though \citet{Branch04} claim
that a mixture of \ion{Fe}{II} and \ion{Co}{II} transitions can
account for these parts of the spectrum without the need for oxygen.
Possible detections of \ion{Ti}{II} \citep{Li03:02cx} and \ion{Cr}{II}
\citep{Branch04} have been claimed for SN~2002cx, but it is extremely
difficult to disentangle the individual contributions of each of these
IGEs in \dc.

\begin{figure*}
\includegraphics[width=18.5cm]{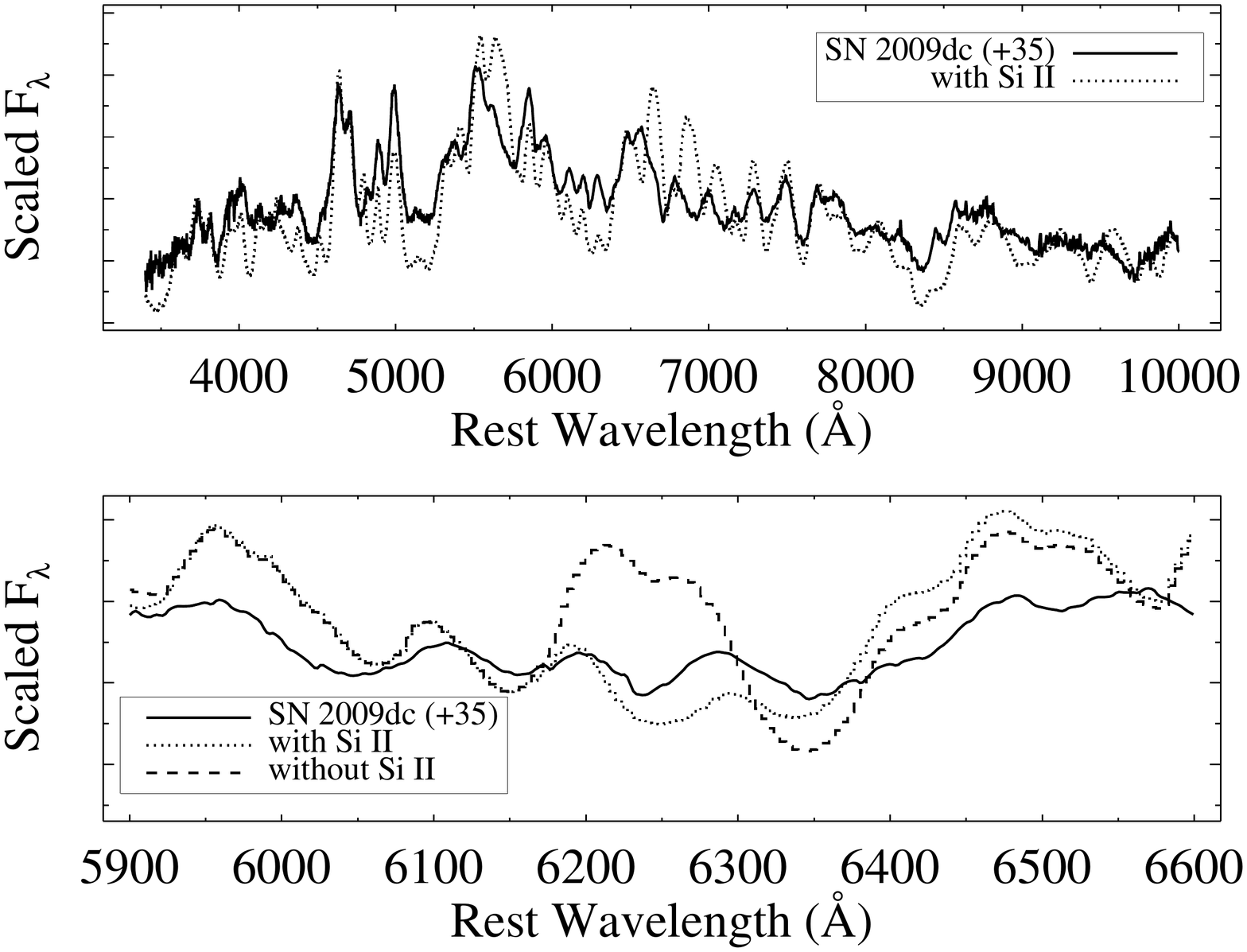}
\caption{The top panel shows our spectrum of \dc\ from 35~days past
  maximum ({\it solid line}) overplotted with our best SYNOW fit which
  includes \ion{Si}{II} ({\it dotted line}). The bottom panel is the
  same as the top panel but zoomed in and centred on the absorption
  feature in \dc\ near 6250~\AA; it also shows the same SYNOW fit but
  with no \ion{Si}{II} ({\it dashed line}).  The SYNOW fit {\it with}
  \ion{Si}{II} appears to reproduce the data better, and thus we
  attribute this absorption to \ion{Si}{II} $\lambda$6355 (blueshifted
  by \about5500~\kms).  We have dereddened our data by
  $E(\bv)_\textrm{MW}=0.070$~mag and $E(\bv)_\textrm{host}=0.1$~mag
  with $R_V = 3.1$. We have also removed the host-galaxy recession
  velocity ($cz = 6300 \pm 300$~\kms) from the
  data.}\label{f:synow_late}
\end{figure*}

The one clear difference between the spectrum of \dc\ and that of
SN~2002cx (and SN~2005hk as well) is that \dc\ has a strong absorption
feature centred at 6250~\AA\ that is not seen in either of these other
peculiar SNe~Ia.  This line is also not present in our spectrum of
\snif\ from a similar epoch.  In Figure~\ref{f:synow_late} we present
a SYNOW fit to our spectrum of \dc\ from 35~days past maximum based
originally on a fit to the spectrum of SN~2002cx from 25~days past
maximum \citep{Li10:02cx}. Both SYNOW fits reproduce the majority of
the strongest features in \dc, and the only difference between the two
is the addition of \ion{Si}{II} at an expansion velocity and
excitation temperature similar to other ions in the fit.  Based on the
comparison of our two fits, we attribute this mysterious absorption in
\dc\ to \ion{Si}{II} $\lambda$6355 (blueshifted by \about5500~\kms).
Note that the spectrum of SN~2005cf from 28~days past maximum, shown
in Figure~\ref{f:one_month_comparison}, has strong \ion{Si}{II}
$\lambda$6355 absorption as well.  The main difference between these
two features is that the absorption in SN~2005cf is blueshifted by
\about9100~\kms, nearly twice the velocity seen in \dc. We discuss
this feature further in \S\ref{sss:latest}.

\subsubsection{2~Months After Maximum}\label{sss:twomonth}

\begin{figure*}
\includegraphics[width=18.5cm]{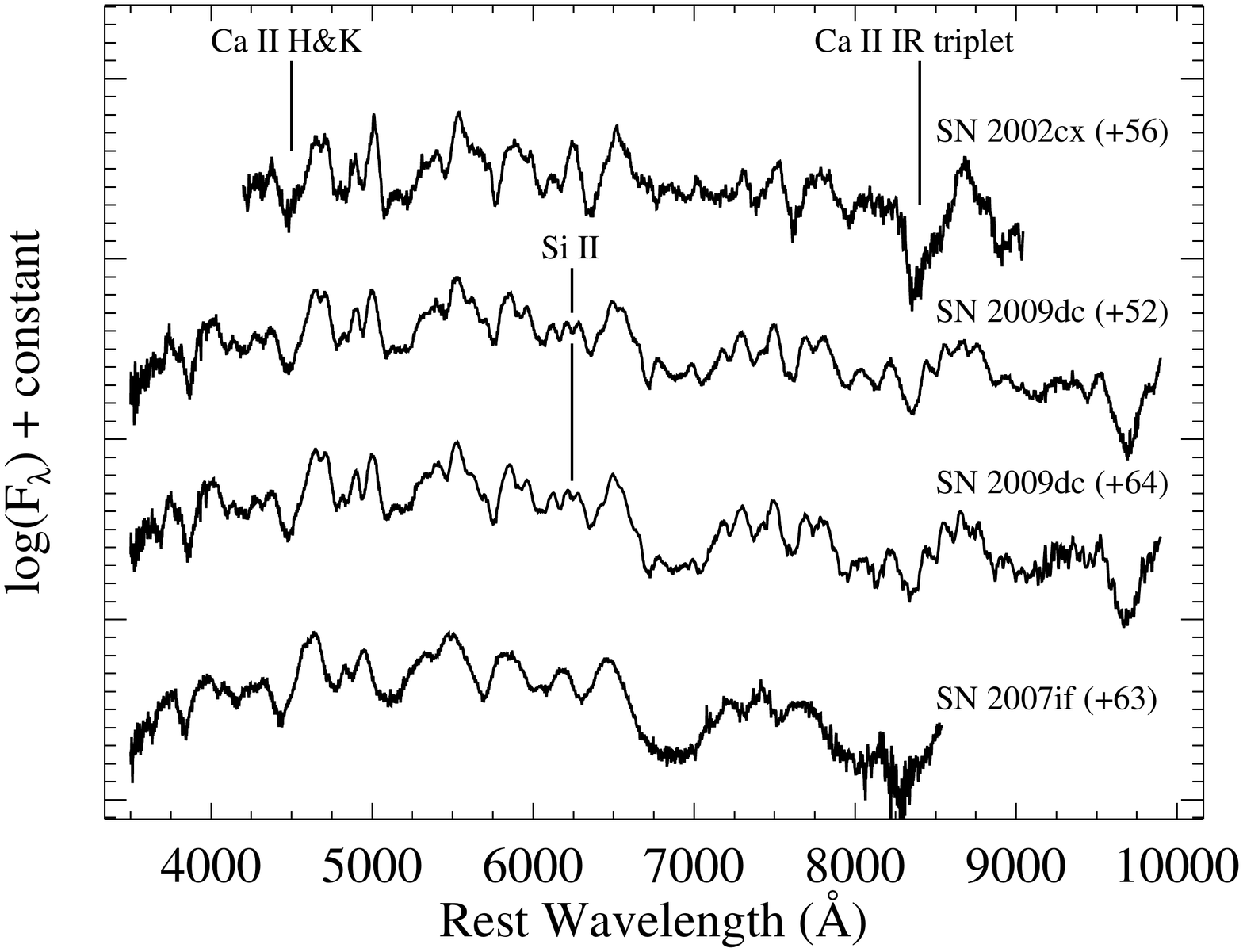}
\caption{Our spectra of \dc\ about 52 and 64~days past maximum, and a
  few comparison SNe with days relative to maximum light indicated
  for each spectrum (in parentheses).  The top SN is the peculiar
  SN~2002cx \citep{Li03:02cx}, the next two spectra are \dc\ (this work),
  and the bottom spectrum is another possible SC~SN~Ia
  \snif\ \citep[][though the displayed spectrum is from our own
    database]{Scalzo10}. Important features discussed in the text are
  labeled. All spectra have had their host-galaxy recession velocities
  removed and have been dereddened according to the values presented
  in their respective references.}\label{f:two_month_comparison}
\end{figure*}

Our day 52 (2009~Jun.~17.5) and day 64 (2009~Jun.~29.3) spectra of
\dc\ are shown in Figure~\ref{f:two_month_comparison}, along with 
spectra of SN~2002cx \citep{Li03:02cx} and \snif\ \citep[][but the
displayed spectrum is from our own database]{Scalzo10}.  Note that \dc\
does not evolve much spectroscopically between days 52 and 64, though
there are a few differences which will be discussed in \S\ref{sss:latest}.

As was the case a month earlier, the spectrum of \dc\ at 2~months past
maximum matches that of SN~2002cx at a similar epoch.  Again, the
overall shape and line profiles look extremely similar between these
two objects for the most part, and both look similar to \snif\ as well
as SN~2005hk \citep{Phillips07}.  At this epoch there remain a handful
of medium-width features in \dc\ that are not seen in \snif, and once
again the velocities of some features in \snif\ appear to be larger
than those of \dc\ by about 1000 \kms.



As mentioned above, the \ion{Si}{II} $\lambda$6355 absorption that was seen in
our day 35 spectrum is still present in our day 52 and day 64 spectra
of \dc.  However, it decreases in strength significantly during these
three observations and disappears almost completely by 80~days past
maximum (see \S\ref{sss:latest}).
Another major difference between \dc\ and SN~2002cx is that
even though the \ion{Ca}{II}~H\&K features are similar in strength,
emission from the \ion{Ca}{II} near-IR triplet is weaker in \dc\ and 
has a complex profile with multiple peaks. We discuss this
feature further in the following section.

\subsubsection{3--4~Months After Maximum}\label{sss:latest}

Figure~\ref{f:post-max} shows all of our post-maximum spectra of
\dc. Overall, the spectra do not change very much from 35 to 109~days
past maximum (see \S\ref{sss:latester} for more information on our
spectrum obtained 281~days past maximum).  Most of the major spectral
features remain constant during these epochs (except for a few notable
examples discussed below) and their profile shapes are largely
unaltered as well. The profiles of the features that do change with
time mainly seem to weaken in a symmetric manner.  That is, there is
no compelling evidence for a preferential increase or decrease of flux
in the blue or red wings of any feature.

One of the most striking aspects of the spectral evolution of \dc\
over these epochs is the disappearance of the broad emission
near 5400~\AA\ and in
the ranges 6500--7200~\AA, 7800--8300~\AA, and (to a lesser extent)
8800--9200~\AA.  The first two of the ranges mainly fall into the $R$
and $I$ bands, respectively.  The emission from these features, which is 
strong through about 1~month past maximum and then begins
to decrease with time, is possibly responsible for the plateau-like
broad peak in the $R$ and $I$ bands near maximum brightness and their
slow late-time decline (as compared to more normal SNe~Ia, see 
Fig.~\ref{figure:phot}). 

\citet{Li03:02cx} also saw these features in SN~2002cx, and their temporal
evolution and the
effect they had on the near-maximum $R$ and $I$-band 
light curves is nearly identical to what we find in \dc.
However, \citet{Li03:02cx} were unable to determine which ions
were responsible for some of the features in these ranges in
SN~2002cx,  
and we are likewise unsure of their origin.  The detailed spectral
modeling required to definitively 
determine their identity is beyond the scope of this paper.

As noted previously, we detect strong \ion{Si}{II} $\lambda$6355
absorption in our day 35 spectrum of \dc.  This feature is still
present, but weakens, in our spectra from 52 and 64~days past
maximum. In our observation of \dc\ at 80~days past maximum there
remains a hint of absorption from \ion{Si}{II} $\lambda$6355, but it
is almost nonexistent and it is effectively undetected in any of our
later spectra.  It is unusual to see such strong absorption from
\ion{Si}{II} $\lambda$6355 in a SN~Ia at such late times.  For
example, at 46~days past maximum the spectrum of SN~2005cf has only a
barely perceptible \ion{Si}{II} $\lambda$6355 feature
\citep[][Fig.~17]{Wang09:05cf}.

The relative strength of this \ion{Si}{II} line at late times in
\dc\ is possibly due to its large ejecta mass. In \S\ref{ss:nickel} we
find that \dc\ has one of the largest \nic\ masses ever produced by a
SN~Ia, and thus one of the largest total ejecta masses as well.  The
massive ejecta may imply that a greater than average amount of silicon
was synthesised in the explosion of \dc, which could give rise to the
rather long-lived \ion{Si}{II} $\lambda$6355 line that we detect.
However, it is also possible that the Si-rich ejecta have an
aspherical or clumpy distribution, and thus the longevity of the
observed \ion{Si}{II} $\lambda$6355 line is simply a viewing angle
effect.

As stated earlier, another major difference between \dc\ and
SN~2002cx is that emission from the \ion{Ca}{II} near-IR triplet is 
weaker in \dc\ and has a complex profile with multiple peaks, even
though the \ion{Ca}{II}~H\&K features are similar in strength.
A SYNOW fit indicates that we are actually resolving the individual
components of the \ion{Ca}{II} near-IR triplet along with shallow
absorption possibly from \ion{Mg}{I} and another, unidentified ion.
The \ion{Ca}{II} and \ion{Mg}{I} features are blueshifted by
\about6000~\kms, which matches fairly well with our previously
calculated post-maximum expansion velocities (see
\S\ref{sss:twomonth}). Thus, it seems that the complex structure seen
in our post-maximum spectra of \dc\ near 8600~\AA\ is a mixture of the
\ion{Ca}{II} near-IR triplet, \ion{Mg}{I}, and probably another ion
whose identity we are unable to definitively determine. 


\subsubsection{9~Months After Maximum}\label{sss:latester}

\begin{figure*}
\includegraphics[width=18.5cm]{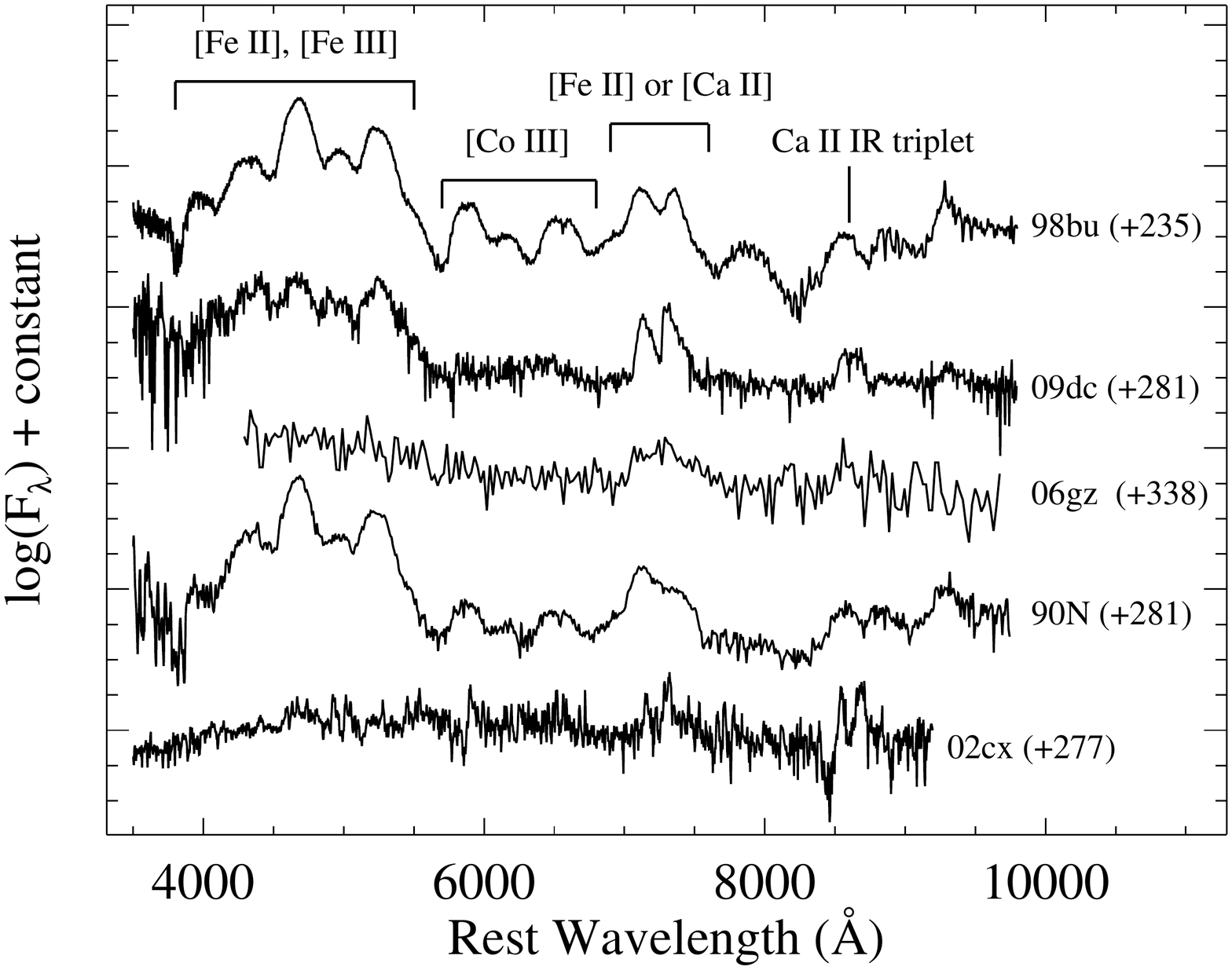}
\caption{Our spectrum of \dc\ about 281~days past maximum, and a few
  comparison SNe with days relative to maximum light indicated for
  each spectrum (in parentheses).  The spectra from top to bottom are
  the normal Type~Ia SN~1998bu \citep{Jha06}, \dc, SN~2006gz
  \citep{Maeda09}, SN~1990N (via SUSPECT), and the peculiar Type~Ia
  SN~2002cx \citep{Jha06}.  Major spectral features are labeled. All
  spectra have had their host-galaxy recession velocities removed and
  have been dereddened according to the values presented in their
  respective references.}\label{f:nine_month_comparison}
\end{figure*}

On 2010~Feb.~6.6 (281~days after maximum), we obtained our final
spectrum of \dc, shown in Figure~\ref{f:nine_month_comparison}
along with late-time spectra of the normal Type~Ia
SN~1998bu \citep{Jha06}, SN~2006gz \citep{Maeda09}, another normal
Type~Ia SN~1990N \citep[][via the online SUSPECT
database\footnote{http://suspect.nhn.ou.edu/$\sim$suspect}]{Gomez98},  
and SN~2002cx \citep{Jha06}. 

At over 9~months past maximum light, \dc\ appears to have completely
transitioned to the nebular phase.  The SN ejecta are now optically
thin and the spectrum shows various emission features.  The broad
peaks in the range 3800--5500~\AA\ are usually attributed to blends of
various features from [\ion{Fe}{II}] and [\ion{Fe}{III}]
\citep[e.g.,][]{Mazzali98}.  These blends appear slightly weaker in
\dc\ as compared to the normal Type~Ia SNe~1998bu and 1990N at 
similar epochs. Interestingly, \citet{Maeda09} found that \gz\
completely lacks these features, though their spectrum has a low S/N. 

The \ion{Ca}{II}
near-IR triplet is detected in both of the normal SNe~Ia presented in
Figure~\ref{f:nine_month_comparison} as well as in \dc\ and possibly \gz\
(though the spectrum is noisy at these wavelengths).
There is essentially no emission in either \dc\ or \gz\ from
\about5700~\AA\ to \about6800~\AA, a region in which [\ion{Co}{III}]
emission has been identified in normal SNe~Ia
\citep{Kuchner94,Ruiz95}.  The ratio 
of emission from the strongest [\ion{Fe}{III}] feature (4589--4805~\AA)
to emission from the strongest [\ion{Co}{III}] feature (5801--5995~\AA)
has been measured in a handful of SNe~Ia at late times; it increases
with both phase relative to maximum and ejecta temperature
\citep{Kuchner94}. 

Using our spectra of SNe~1998bu and 1990N shown in
Figure~\ref{f:nine_month_comparison}, we calculate ratios of about 5.6
and 7.7, respectively.  These fit nicely with the other data displayed
in Figure~2 of \citet[][]{Kuchner94}.  However, it is nearly
impossible to calculate this ratio for \dc\ and \gz, since they appear
to lack almost any putative cobalt emission at this epoch (and the
cobalt and iron lines in spectra at earlier epochs are still
significantly blended).  We attempt to measure the tiny amount of
[\ion{Co}{III}] emission above the background level from 5801~\AA\ to
5995~\AA\ and calculate a Fe/Co ratio of about 27.  This value, which
is likely an underestimate, is already well above that of any other
SN~Ia at this phase presented by \citet{Kuchner94}.  It is also above
the dotted lines in Figure~2 of \citet[][]{Kuchner94}, implying that
the ejecta temperature of \dc\ may be $\gtrsim10,000$~K. Even though
this is broadly consistent with our relatively large value of 20,000~K
for the blackbody temperature of \dc\ before maximum
(\S\ref{sss:synow}), 10,000~K is extremely hot for a normal SN~Ia
9~months past maximum brightness and the shape of the spectrum at this
epoch does not seem to support such a high temperature.

The most well-defined spectral feature in our day 281 spectrum of
\dc\ is the broad, double-peaked line centred near 7200~\AA, which was
also the strongest feature detected in a spectrum of \gz\ from
338~days past maximum \citep{Maeda09}.  In late-time spectra of SNe~Ia
this is attributed to blends of various [\ion{Fe}{II}] lines, whereas
in core-collapse SNe it is associated with [\ion{Ca}{II}]
$\lambda\lambda$7291, 7324, though in a late-time spectrum of \gz\ it
has been suggested that the feature is due to a combination of both
[\ion{Fe}{II}] and [\ion{Ca}{II}] lines \citep{Maeda09}.  The shape of
the profile in \dc\ matches roughly that of SN~1998bu (though the
central dip in \dc\ is deeper), but differs somewhat significantly
from that of SN~1990N.  Shapes similar to the one seen in \dc\ have,
however, been seen by \citet{Maeda09}, who present synthetic late-time
spectra of SNe~Ia.  Their models with more massive progenitors seem to
lead to the feature near 7200~\AA\ becoming strong relative to the
complex of lines in the range 3800--5500~\AA, as seen in \dc.

\subsubsection{SN~2009dc and SN~2002cx-Like Objects}\label{sss:cx}

As already mentioned numerous times, \dc\ shares some interesting
properties with both SN~2002cx and SN~2005hk, which are members of a 
class of peculiar SNe~Ia \citep[][]{Filippenko03,Li03:02cx,Jha06}.
These so-called ``SN~2002cx-like'' SNe are characterized by low
expansion velocities and low peak luminosities, slow late-time declines in
$R$, and \ion{Fe}{III} features dominating early-time spectra 
\citep{Jha06}. Many of these criteria hold for
\dc\ as well, but the very obvious difference is that the peak
luminosity of \dc\ is much larger than average (see \S\ref{ss:lum})
while those of SN~2002cx-like objects are well below average.

In \S\ref{ss:lc}, we noted the absence of a prominent second
maximum in the $R$ and $I$-band light curves of \dc.  This lack of a
secondary maximum has been observed in SN~2002cx-like objects 
\citep{Li03:02cx,Phillips07}. However, as seen in Figure~\ref{figure:phot},
SN~2005hk (a SN~2002cx-like object) evolves much faster than \dc\ in
all bands and proves to be a poor match photometrically.  The colour indices
of \dc\ and SN~2005hk are again a poor match as shown in
Figure~\ref{figure:color}.  

Spectroscopically, \dc\ does not really resemble SN~2002cx-like
objects before maximum (see Fig.~\ref{f:early_spec}), nor about
9~months after maximum (see Fig.~\ref{f:nine_month_comparison}).
At these latest times, the spectrum of \dc\ consists of broad
emission from forbidden iron lines while the spectrum of SN~2002cx is
made up of many {\it narrow permitted} lines of 
\ion{Fe}{II}, \ion{Na}{II}, \ion{Ca}{II}, and (tentatively) \ion{O}{I}
\citep{Jha06}. This is supported by the fact that the narrow spectral
features of SN~2002cx do not simply appear to correspond to resolved
versions of the blends seen in \dc. 

Despite all of these differences, we have shown that at 1--2~months
past maximum brightness the spectra of \dc\ and SN~2002cx are
surprisingly similar.  In Figure~\ref{f:one_month_comparison} we have
combined the day 20 and 25 spectra of SN~2002cx from \citet{Li03:02cx}
in order to improve the S/N and the wavelength coverage.  Not only are
the overall shapes of these two spectra nearly identical, but the
width, strength, and existence of almost all of the individual
spectral features match extremely well (with one significant
exception, the \ion{Si}{II} $\lambda$6355 absorption as mentioned in
\S\ref{sss:onemonth}).  Both objects, after being dereddened and
deredshifted, appear to have an underlying blackbody continuum of
about 5500~K \citep{Branch04}.  

About a month past maximum, \citet{Li03:02cx} and \citet{Branch04}
calculate expansion velocities of \about 5000--6000~\kms\ for
SN~2002cx, and \citet{Phillips07} determine expansion velocities in
nearly the same range for SN~2005hk.  For \dc\ we also derive
velocities in this range for a few ions (\ion{O}{I}, \ion{Ca}{II},
\ion{Fe}{II}, and \ion{Si}{II}) with significant uncertainty since not
only are many of these features blended, but the line identifications
themselves are not definitive.  These values also match well with the
expansion velocity of \about6000~\kms\ derived by \citet{Yamanaka09}
from their spectrum of \dc\ from 25~days past maximum.

Both \citet{Li03:02cx} and \citet{Branch04} determine an expansion
velocity for SN~2002cx at 56~days past maximum of
\about4700~\kms\ based on the \ion{Fe}{II} $\lambda$4555 feature and a
SYNOW fit.  \citet{Phillips07} calculate a slightly higher value of
about 6000~\kms\ for SN~2005hk at 55~days past maximum (also using the
\ion{Fe}{II} $\lambda$4555 line). Based on this \ion{Fe}{II} feature,
we determine an expansion velocity of \about5500~\kms\ for our spectra
of \dc\ from 52 and 64~days past maximum.  If we use the \ion{Si}{II}
$\lambda$6355 line, which is still present in \dc\ but not seen in the
SN~2002cx-like objects, we derive similar expansion velocities for
both our day 52 and day 64 spectra. In general, it appears that the
post-maximum expansion velocities of the spectral features of \dc\ are
slightly greater than those seen in SN~2002cx by a few hundred
\kms\ depending on the feature \citep{Li03:02cx,Branch04}, whereas they
are about equal to those of SN~2005hk \citep{Phillips07}.

In conclusion, \dc\ and SN~2002cx-like objects have quite different
photometric properties but are nearly identical spectroscopically {\it
  only} at 1--2~months past maximum brightness.  The expansion
velocity, the blackbody temperature, and the chemical composition
match extremely well at these epochs, but the evolution of these
observables from pre-maximum to nearly a year after maximum is quite
different between \dc\ and SN~2002cx-like objects. 

\subsubsection{SN~2007if}\label{sss:if}
\begin{figure*}
\includegraphics[width=18.5cm]{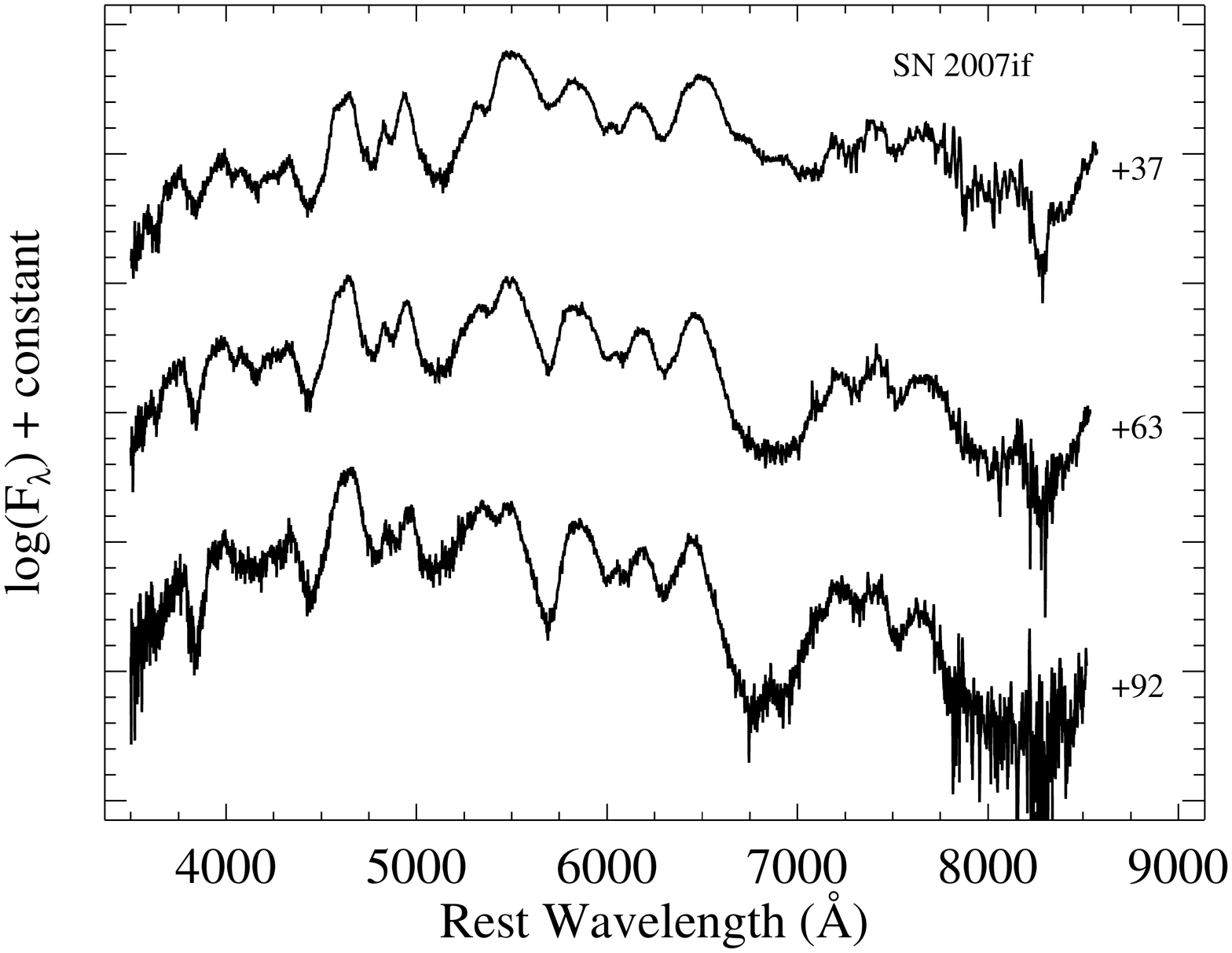}
\caption{Our spectra of \snif\ with days relative to maximum light
  indicated in each case. We have removed the host-galaxy recession
  velocity \citep[$cz =$ 22,200 $\pm
    250$~\kms,][]{Scalzo10}.}\label{f:07if}
\end{figure*}

In Figure~\ref{f:07if} we present our late-time spectra of another
possible SC~SN~Ia, \snif.  As noted earlier, the spectra of
\snif\ generally look similar to those of \dc, SN~2002cx, and even
SN~2005cf at comparable epochs.  The most significant spectral
differences between \snif\ and \dc\ are, as stated previously, that
the expansion velocities seen in \snif\ are \about1000~\kms\ larger
than those in \dc\ and SN~2002cx at comparable phases.  This larger
velocity is likely causing some of the medium-width features seen in
\dc\ and SN~2002cx to be ``smoothed out'', and hence we do not detect
many of these features in \snif.  It also has velocities which are
smaller than those seen in SN~2005cf, but otherwise the spectra match
well.

We have also stated previously that \snif\ does not show the strong
\ion{Si}{II} $\lambda$6355 absorption after maximum that is visible in
spectra of \dc\ (as well as in SN~2005cf).  Unfortunately, our
observations of \snif\ do not extend to sufficiently long wavelengths
for us to inspect whether the \ion{Ca}{II} near-IR triplet of \snif\ 
shows the complex structure seen in \dc.

The main aspects of the spectral evolution of \snif\
over our three epochs is the disappearance of broad
emission near 5500~\AA\ and in
the ranges 6700--7100~\AA\ and possibly 7800--8100~\AA\ (though our
data are noisy in this second range of wavelengths).  These match nearly exactly
with the spectral features of \dc\ that were seen to decrease in
strength over similar epochs.\footnote{The third wavelength range
  which possibly decreased in strength in \dc, 7800--8100~\AA, is
  outside of our spectral coverage of \snif.} 

\subsection{Reddening and Extinction}\label{ss:reddening}

As mentioned above, we detect weak (but noticeable), narrow absorption
from \ion{Na}{I}~D in all of our spectra of \dc, both at zero redshift
and at the recession velocity of the SN itself ($cz = 6300 \pm
300$~\kms). Even though there are clear notches in all of our spectra
at the correct wavelengths for \ion{Na}{I}~D in the MW and in the host
of \dc\ (UGC~10064), they are not broader than our spectral
resolution.

We attempt to measure an EW of both absorption features in each of our
observations, and we calculate average values of \about$0.7 \pm
0.1$~\AA\ for the MW and \about$1.0 \pm 0.1$~\AA\ for UGC~10064. These
are close to the values of 0.7~\AA\ and 1.2~\AA\ reported by
\citet{Marion09} and 0.5~\AA\ and 1.0~\AA\ found by \citet{Tanaka09}.

There is some uncertainty in how exactly to convert the EW of
\ion{Na}{I}~D absorption into extinction, but a few (mainly empirical)
relations do exist \citep[e.g.,][]{Barbon90,Munari97,Turatto03}.  Using
these conversions, we determine a range of values for $E(\bv)$ both
in the MW ($0.1\lesssim E(\bv) _\textrm{MW} \lesssim 0.18$~mag) and in 
UGC~10064 ($0.15\lesssim E(\bv) _\textrm{host} \lesssim 0.3$~mag).

For the extinction from the MW, we compare our calculated values of
$E(\bv)$ to the value from the Galactic dust maps of
\citet{Schlegel98}.  Their maps give $E(\bv)_\textrm{MW} = 0.070$~mag
for the position of \dc, slightly lower than our range of values.
However, since the accuracy of the \citet{Schlegel98} dust maps is
almost certainly higher than that of our EW measurements, we adopt
$E(\bv) _\textrm{MW} = 0.070$~mag for our analysis.

The extinction from UGC~10064 is a bit more complicated to determine
accurately. If the Lira-Phillips relation holds for \dc, then we
expect an extinction of $E(\bv) _\textrm{host} \approx 0.26$~mag (see
\S\ref{ss:cc} for details on our application of the Lira-Phillips
relation), which is within our range of values calculated from the
spectra. Furthermore, if we assume that the EW of \ion{Na}{I}~D is
directly proportional to $E(\bv)$, as \citet{Tanaka09} do, then we
derive $E(\bv)_\textrm{host} \approx \left( 1.0\textrm{ \AA
}/0.7\textrm{ \AA}\right)\times0.070\approx0.1$~mag.  Taking into
account all of these values, their uncertainties, and the method by
which they were derived, we adopt $E(\bv)_\textrm{host} = 0.1$~mag for
our analysis, which is our lowest plausible, nonzero
reddening. However, we note that the reddening could be as large as
0.3~mag.

\subsection{Host-Galaxy Properties}\label{ss:host}

\dc\ occurred \about27\arcsec\ away from the centre of the S0 galaxy
UGC~10064.  We adopt the luminosity distance ($d_{L}$) as the distance
to the SN, again assuming a $\Lambda$CDM Universe with $H_{0} =
70$~\kms~\perMpc, $\Omega_{m} = 0.27$, and $\Omega_{\Lambda} = 0.73$
\citep{spergel07} and $z = 0.022 \pm 0.001$, where the redshift has
been corrected for infall into the Virgo cluster and we have adopted
an uncertainty of $cz =300$~\kms\ to allow for a peculiar velocity
induced by the gravitational interaction of neighboring galaxies. At
$d_{L} = 97 \pm 7$~Mpc, \dc\ is a projected distance of \about13~kpc
from the centre of UGC~10064.  A useful galactic scale is the radius
containing 50\% of the flux \citep{Petrosian76}, which for UGC~10064
is about $4\farcs71 \pm 0\farcs21$ in the $R$ band
\citep{Abazajian09}.  This puts \dc\ about 5.7 Petrosian radii (in
projection) from the centre of its host.  For comparison, \fg\ was
found to be projected 0.9~kpc from the core of its low-mass
star-forming host \citep{Howell06} and \snif\ was discovered
effectively directly on top of its low-mass and low-metallicity host
\citep{Scalzo10}.  The host of \gz\ is an Scd galaxy, and the SN was
projected 14.4~kpc from the nucleus \citep{Hicken07}, while SN~2004gu
was projected 2.21~kpc from its host's nucleus \citep{Monard04}.  It
seems that host-galaxy type and distance from the nucleus vary widely
among candidate SC~SN~Ia (though the distance comparison is difficult
given projection effects).  However, \fg, \snif, and possibly
\dc\ (see below) were all associated with regions of recent star
formation.

\citet{Wegner08} inspected an optical spectrum of the core of
UGC~10064 and derived an age of $4.0 \pm 3.0$~Gyr, a metallicity
[$Z$/H] $ = 0.45  
\pm 0.07$, and a chemical abundance [$\alpha$/Fe] $ = 0.13 \pm 0.09$.  
This value for metallicity is about 1$\sigma$ above the average
found for 26 S0/E void galaxies by \citet{Wegner08}, whereas the
chemical abundance is just about the same as their average value.
Optical spectra of the nucleus of UGC~10064 show no strong
emission lines \citep{Wegner08,Abazajian09}, implying very little
ongoing star formation ($\lesssim0.2$~\msun\ yr$^{-1}$) in the core of
the galaxy. 

However, recent deep \ion{H}{I} surveys of S0 galaxies have shown that the
majority of noncluster early-type galaxies contain substantial
reservoirs ($10^6$--$10^9$~\msun) of neutral gas that extend to large
galactocentric radii \citep{Morganti06}, whose morphology and
kinematics suggest external origins.  
On the other hand, in the majority of these 
galaxies the column density of this gas is too low to support star
formation at large radii, and thus the stellar populations
throughout the galaxy are old.  

Interestingly, in a fraction of the early-type galaxies rich in
\ion{H}{I}, a dynamical feature such as a prominent bar can create
relatively dense ring-like accumulations of neutral gas at large radii
\citep[e.g.,][]{Schinnerer02,Weijmans08}.  In these regions of
enhanced \ion{H}{I}, low-level star formation
(\about0.1~\msun\ yr$^{-1}$) is observed \citep{Jeong07,Shapiro09}.
Since \dc\ exploded in the outskirts of an S0 galaxy, it seems
difficult to say with certainty whether \dc\ came from an old or young
stellar population.

In the Supplementary Information of \citet{Howell06}, the age of the
host galaxy of \fg\ is estimated to be \about700~Myr (with significant
uncertainty) and the star-formation rate to be
$1.26^{+0.02}_{-1}$~\msun\ yr$^{-1}$, averaged over 0.5~Gyr.  This
star-formation rate is nearly an order of magnitude greater that the
value derived by \citet{Wegner08} for the core of UGC~10064 as well as
the residual star-formation rate seen in rings of enhanced neutral gas
at large galactocentric radii \citep{Jeong07,Shapiro09}.  SC~SNe~Ia
are expected to be found preferentially in young stellar populations
\citep[e.g.,][and references therein]{Howell06,Chen09}, but they are
not necessarily restricted to them.  If SC~SNe~Ia are the result of
the merger of two degenerate objects, then the merger timescale can be
set by the emission of gravitational waves \citep[e.g.,][]{Iben84}.
Depending on the particular orbital parameters of the binary, there
may be a significant delay between the formation of the constituent
degenerate objects and their eventual merger.  Thus, SC~SNe~Ia {\it
  could} occur in older stellar populations.

Even though there is a significant difference between the average
star-formation rates near the sites of \dc\ and \fg, there is still a
possibility that they both come from relatively young stellar
populations.  As mentioned above, an age of \about700~Myr was
calculated for the environment of \fg\ \citep[][Supplementary
  Information]{Howell06}, while an age of \about4.0~Gyr was calculated
for the {\it core} of the host of \dc\ \citep{Wegner08}.  The fact
that \dc\ is a few effective radii away from the nucleus of its host,
coupled with the observations of low levels of residual star formation
in other early-type galaxies at such galactocentric radii, imply that
it is possible that \dc\ also came from a relatively small, but young,
stellar population on the edge of UGC~10064, and that the age
calculated for the core of the galaxy is not actually representative
of the local environment of \dc.

\begin{figure}
\includegraphics[width=8.5cm]{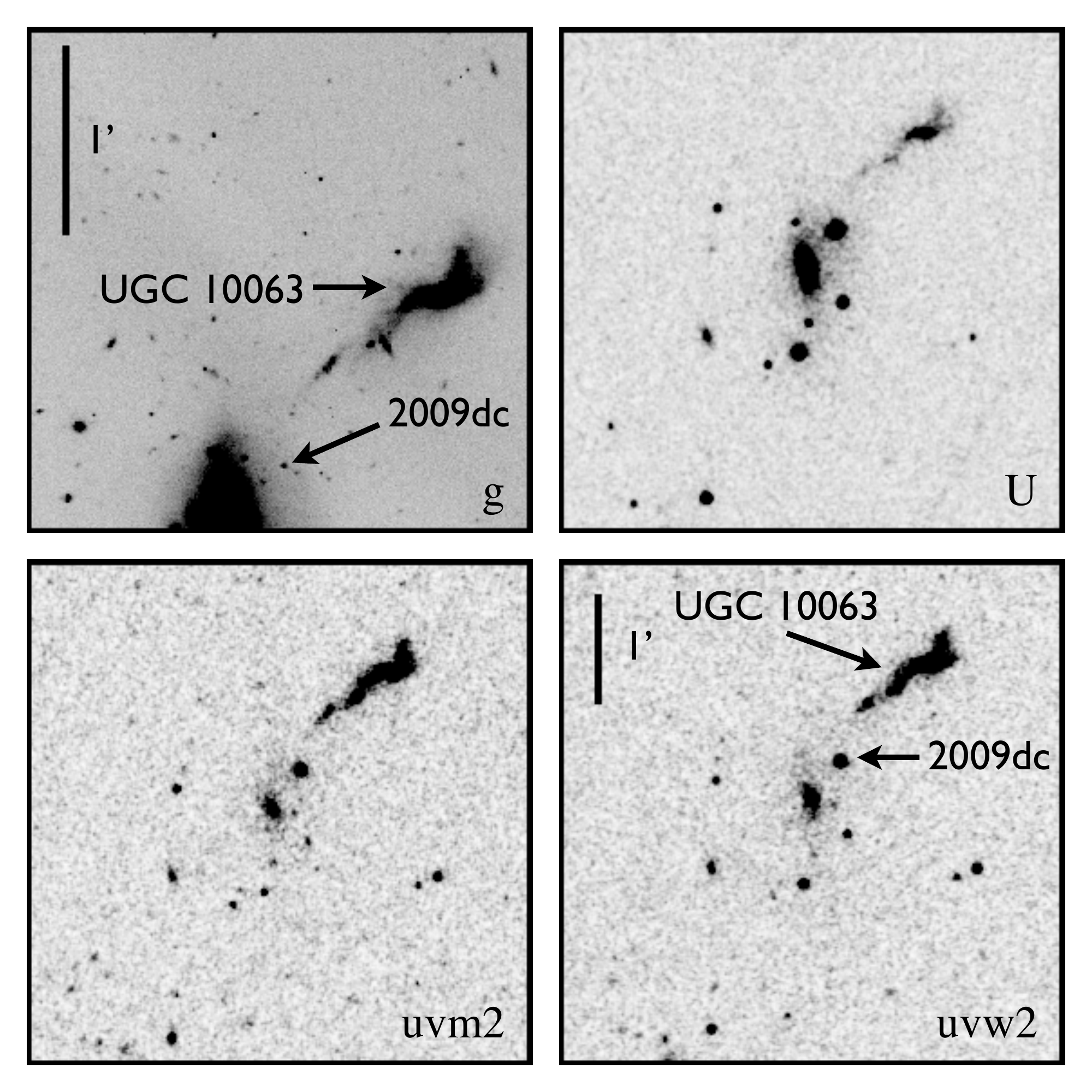}
\caption{Keck $g$-band image and stacked UVOT $U$, $uvm2$, and $uvw2$
  images of UGC~10063 (SDSS~J155108.42+254321.3).  UGC~10063 and
  \dc\ are labeled in the upper-left and lower-right panels. The
  upper-left panel is 2.69\arcmin$\times$2.69\arcmin, and the
  remaining panels are 4.67\arcmin$\times$4.67 \arcmin; north is up
  and east is to the left in all four images.  UGC~10063 increases
  in prominence in progressively bluer bands, and is much bluer than
  UGC~10064 (compare its appearance here with that in
  Fig.~\ref{f:galaxy}).  It also seems to have a tidal tail extending
  toward UGC~10064, pointed roughly in the direction of the site of
  \dc.\label{f:BBB} }
\end{figure}

Furthermore, in the SDSS Data Release 7, an extended, blue source
appears 99$\farcs$3 (which is a projected distance of 43.6~kpc) to the
northwest of UGC~10064 \citep{Abazajian09}.  This SBdm galaxy is known
as SDSS~J155108.42+254321.3 or UGC~10063 \citep{deVaucouleurs91}.
{\it Swift}/UVOT images of the \dc\ field (shown in Fig.~\ref{f:BBB})
indicate that UGC~10063 is significantly bluer than UGC~10064. The
irregularly shaped UGC~10063 is barely visible in the KAIT $R$-band
image to the northwest of \dc\ in Figure~\ref{f:galaxy}. Stacked UVOT
images show that UGC~10063 increases in prominence in progressively
bluer bands. As seen in Figure~\ref{f:BBB}, UGC~10063 seems to have a
tidal tail extending toward UGC~10064, pointed roughly at the site of
\dc.

\begin{figure*}
\includegraphics[width=18.5cm]{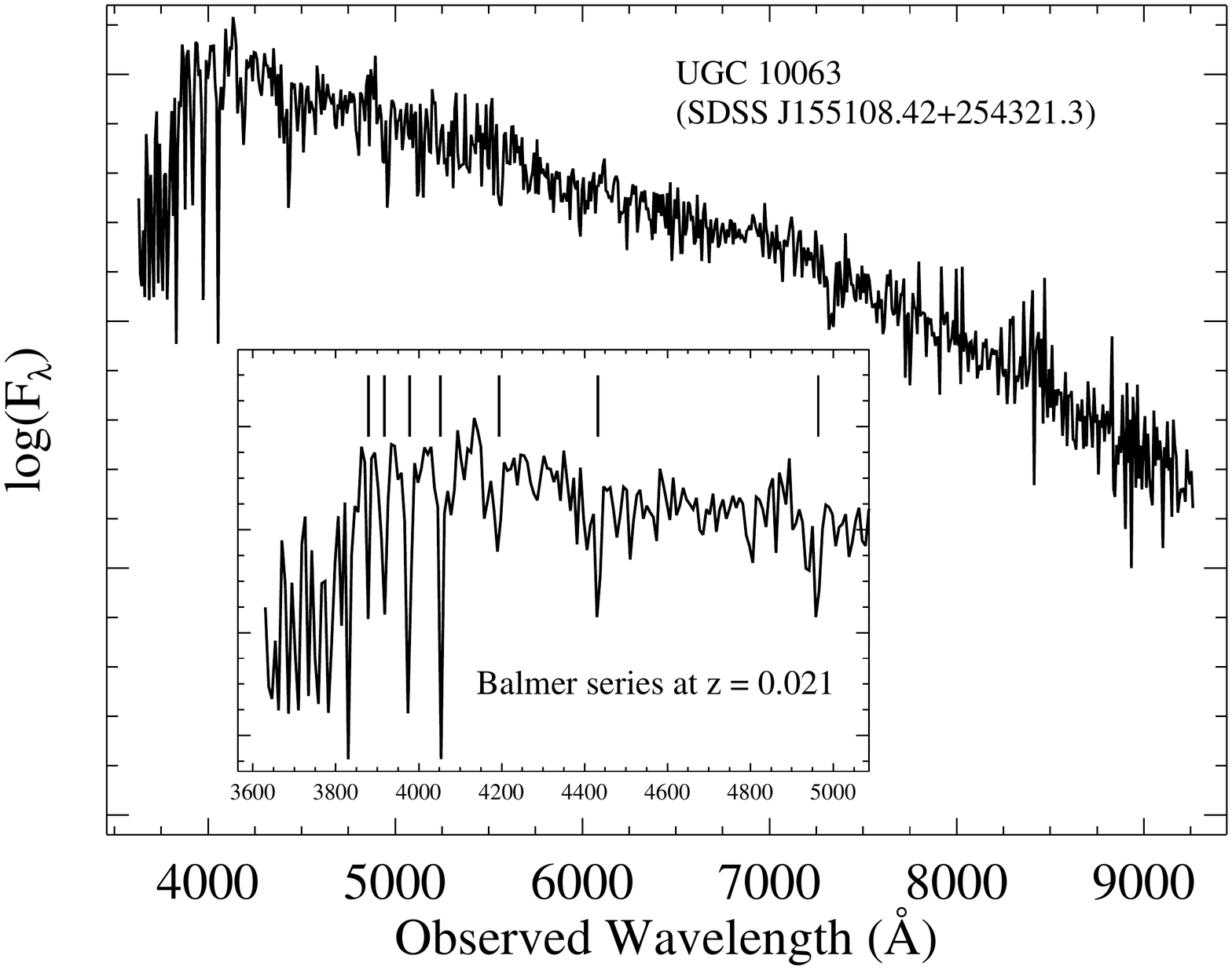}
\caption{An optical spectrum of UGC~10063
  (SDSS~J155108.42+254321.3). The inset is a zoom-in of the full
  spectrum.  Narrow Balmer-series absorption (\hbeta\ through
  H$\theta$) is marked, and these features yield $z = 0.021 \pm
  0.001$, equal to the redshift of UGC~10064 and \dc. Note that
  \halpha\ is undetected at an expected wavelength of \about6700~\AA;
  perhaps \halpha\ emission from residual \ion{H}{II} regions fills in
  the absorption line.}\label{f:BBB_spec}
\end{figure*}

On 2010~Feb.~7 we obtained a spectrum of UGC~10063 using LRIS on the
Keck~I telescope (with a $1''$-wide slit, a 600/4000 grism on the blue
side, and a 400/8500 grating on the red side, resulting in FWHM
resolutions of \about4~\AA\ and \about6~\AA, respectively).  It has a
blue continuum with few spectral features, but there is strong and
narrow absorption from the Balmer series (though \halpha\ is
undetected).  Strong Balmer lines and the lack of obvious emission
lines are the hallmarks of so-called ``post-starburst'' galaxies in
which vigorous star-formation activity ended between \about50~Myr and
1.5~Gyr ago \citep[][and references therein]{Poggianti09}. This is
consistent with UGC~10063 being a SBdm galaxy \citep{deVaucouleurs91}.
From the seven Balmer absorption lines marked in
Figure~\ref{f:BBB_spec}, we calculate the redshift of UGC~10063 to be
$0.021 \pm 0.001$, consistent with the redshift derived from the 21-cm
\ion{H}{I} line \citep[$z = 0.02158 \pm
  0.00003$,][]{Schneider90}. This also exactly matches the redshift of
UGC~10064 and \dc\ (as determined from narrow \ion{Ca}{II} H\&K
absorption).

Based on the spectrum of UGC~10063 and its position relative to
UGC~10064, we propose that these two galaxies had a close encounter in
the relatively recent past.  If UGC~10063 passed near the nucleus of
UGC~10064, gas from UGC~10063 could have been stripped off and become
gravitationally bound to UGC~10064 \citep[which has been seen in both
simulations and observations, e.g.,][]{Barnes91,Barroso04}.  This
newly acquired gas (some of which could be at large galactocentric
radii) would have likely undergone a burst of star formation soon
after the two galaxies had their closest encounter
\citep{Mihos94,Barnes96}, and this might be how and when the
progenitor of \dc\ formed. Both \snif\ and \fg\ were also found in
hosts with recent star formation \citep{Scalzo10}; in fact, \fg\ was
discovered in a tidal feature \citep{Howell06} reminiscent of \dc\ and
UGC~10063.

\section{Discussion}\label{s:discussion}

\subsection{Bolometric Luminosity}\label{ss:lum}


To estimate the bolometric luminosity of \dc, we follow the method
outlined by \citet{Howell09}. Using spectra of \dc\ presented here and
spectra of the similar SC SN~Ia candidate \snif\ from
\citet{Scalzo10}, we take each simultaneous epoch of \bvri\ photometry
and warp the spectrum closest in phase to match the photometry using a
smooth third-order spline.  We then integrate the warped spectrum over
wavelengths of 4000--8800~\AA\ (i.e., from the blue edge of the
$B$-band filter to the red edge of $I$-band filter.)  We convert the
flux into a luminosity using the luminosity distance ($d_{L} = 97 \pm
7$~Mpc) adopted in \S\ref{ss:host}.

The above method will only give us the optical bolometric luminosity.
To derive the total bolometric luminosity from UV to IR, we use
bolometric corrections from \citet{Wang09:05cf}, who found that
\bvri\ only accounts for $\sim 60$\% of the total luminosity in the
case of SN~2005cf, a normal SN~Ia. They also found that an additional
20\% is emitted in the UV and $U$ band, and the final 20\% in the
near-IR.  We test to see if the tabulated corrections for the UV and
$U$ band are reasonable for \dc\ by first calculating the bolometric
luminosity using optical and UV photometry at $t =0$~days (using our
optical ground-based photometry combined with UVOT photometry of
\dc). We find that our results are consistent with a correction of
20\%. \cite{Yamanaka09} found that their near-IR data for
\dc\ represented \about20\% of the bolometric luminosity.  With this,
we cautiously adopt a 40\% correction for our bolometric luminosity
using our well-sampled \bvri\ photometry. Overall, SN~2005cf is a much
different SN than SN~2009dc, so we include a systematic uncertainty of
20\% for our bolometric luminosity.



\begin{figure}
\includegraphics[width=8.5cm]{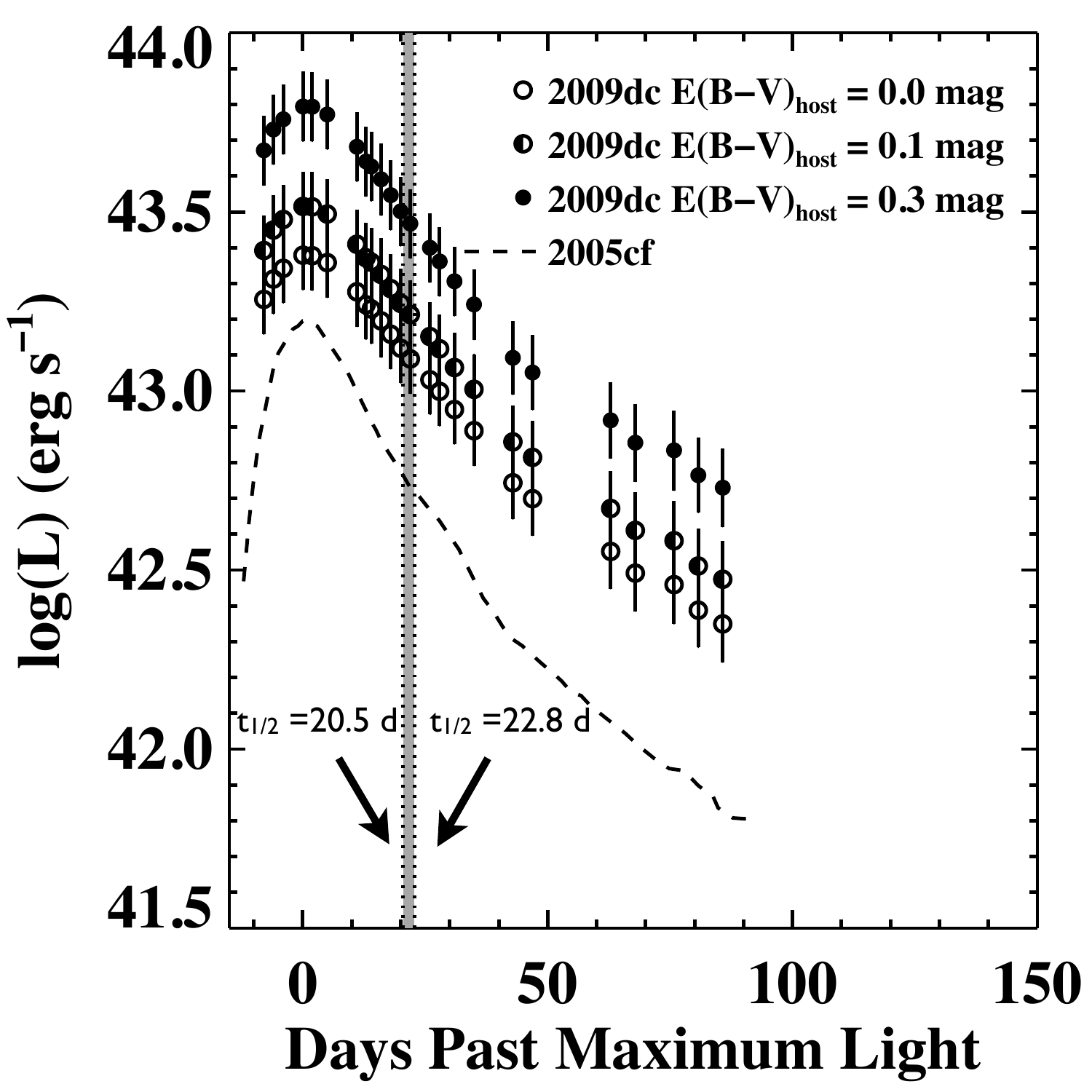}
\caption{Bolometric light curve of \dc\ assuming various levels of
  extinction, in comparison with that of SN~2005cf. The curve plotted
  with empty circles assumes no host-galaxy extinction, the
  half-filled circles assume $E(\bv)_\textrm{host} = 0.1$~mag with
  $R_{V} = 3.1$, and the filled circles assume $E(\bv) _\textrm{host}
  = 0.3$~mag with $R_{V} = 3.1$. For comparison, we include the
  bolometric light curve of SN~2005cf as a dashed line
  \citep{Wang09:05cf}.  We indicate our calculated range of values of
  $t_{+1/2}$ (i.e., the number of days after maximum bolometric
  luminosity when the bolometric luminosity has decreased to half its
  maximum value) for \dc\ within the shaded grey region bounded by
  dotted lines. The lower bound of $t_{+1/2} = 20.5$~days is
  calculated using a host reddening of $E(\bv)_\textrm{host} =
  0.3$~mag and the upper limit of $t_{+1/2} = 22.8$~days assumes no
  host-galaxy extinction.\label{fig:bol_lum}}
\end{figure}

Figure~\ref{fig:bol_lum} shows our final bolometric light curves for
the various levels of host-galaxy extinction we have adopted
throughout this paper. We also plot the bolometric light curve of
SN~2005cf for comparison.


If we use our fiducial nonzero value for the host-galaxy reddening,
$E(\bv)_\textrm{host} = 0.1$~mag, and our largest reasonable value for
the host reddening, $E(\bv)_\textrm{host} = 0.3$~mag, then we get $M_V
= -20.16 \pm 0.1$~mag ($L_\textrm{bol} = (3.3 \pm
0.6)\times10^{43}$~\ergps) and $M_V = -20.82 \pm 0.10$~mag
($L_\textrm{bol} = (7.3 \pm 1.0)\times10^{43}$~\ergps),
respectively. We note that assuming no host-galaxy extinction we find
$M_{V} = -19.83 \pm 0.10$~mag ($L_\textrm{bol} = (2.4 \pm
0.5)\times10^{43}$~\ergps). A summary of these results can be found in
Table~\ref{t:bol_lum}.  Our range of values of $L_\textrm{bol}$ for
\dc\ matches quite well that of \citet{Yamanaka09}:
($2.1$--$3.3)\times10^{43}$~\ergps.  The most significant difference
between the two analyses comes from using a different range of values
for $E(\bv)_\textrm{host}$ and $R_V$.

\begin{table}
\centering
\caption{Bolometric luminosity for various reddenings
\label{t:bol_lum}}
\begin{tabular}{c c c}
\hline
$E(\bv)_\textrm{host}$ &
$L_\textrm{bol,max}$   &
$M_{V,\textrm{max}}$ \\
(mag) & ($10^{43}$~\ergps) &(mag)\\
\hline
0.0 & $2.4 \pm 0.5$ & $-19.83 \pm 0.10$ \\
0.1 & $3.3 \pm 0.6$ & $-20.16 \pm 0.10$ \\
0.3 & $7.3 \pm 1.0$ & $-20.82 \pm 0.10$ \\
\hline
\end{tabular}
\end{table}

\subsection{\nic\ Masses and Energetics}\label{ss:nickel}

\subsubsection{\nic\ Masses from Arnett's Law}\label{sss:arnett}

The vast majority of the bolometric luminosity of SNe~Ia comes from
the decay of \nic\ to $^{56}$Co (and subsequently the decay of
$^{56}$Co to $^{56}$Fe),
so we can calculate how much \nic\ was created in \dc\ by using
``Arnett's law'', which asserts that the luminosity of a SN~Ia at
maximum light is proportional to the instantaneous rate of radioactive
decay of \nic\ \citep[e.g.,][]{Arnett82,Stritzinger05}.
Arnett's law is often written
\begin{equation}
M_\textrm{Ni} = \frac{L_\textrm{bol}}{\alpha \dot{S}(t_R)} ,
\end{equation}
where $M_\textrm{Ni}$ is the mass of \nic\ present in the ejecta,
$L_\textrm{bol}$ is the bolometric luminosity at maximum light,
$\alpha$ is the ratio of bolometric to radioactive luminosities (which
is of order unity), and $\dot{S}(t_R)$ is the radioactive luminosity
per mass of \nic\ as a function of the rise time, $t_R$.

For \dc, we have tight constraints on the rise time from our
photometric data and, as described in \S\ref{ss:lc}, we find a value
of $23 \pm 2$~days.  Substituting this into the equation for
$\dot{S}(t_R)$ from \citet{Stritzinger05}, we get $\dot{S}(t_R)
\approx (1.65 \pm 0.13)\times10^{43}$~\ergps~\msun$^{-1}$.
\citet{Stritzinger05} find $\alpha = 1.2 \pm 0.2$, which we use for
our analysis, though we note that this is an upper limit (which leads
to a lower limit on the \nic\ mass) since any \nic\ that is {\it
  above} the photosphere will not contribute to the SN luminosity.

Using these values, along with our most likely estimate of the
bolometric luminosity mentioned above, we calculate that \dc\
synthesised about $1.7 \pm 0.4$~\msun\ of \nic.
If we adopt the bolometric
luminosity assuming no host-galaxy reddening
and our largest reasonable host-galaxy reddening,
along with our calculated rise time and $\alpha=1.2$, we get
\about1.2~\msun\ and \about3.7~\msun\ of \nic, respectively. 

\subsubsection{\nic\ Masses from Late-Time Photometry}\label{sss:late_time}

The detection of \dc\ in late-time images at levels similar to those
of normal SNe~Ia offers a stark contrast to \gz, another SC~SN~Ia
candidate. Late-time photometry obtained by \cite{Maeda09}
\about360~days after maximum found that \gz\ had faded significantly
faster than normal SNe~Ia, casting doubt on the amount of
\nic\ synthesised in the explosion.  The authors constrain $M_{\rm Ni}
\lesssim 0.3$~\msun\ assuming that the positrons are nearly completely
trapped in the ejecta. This is at odds with the \about1~\msun\ of
\nic\ calculated to power the light curve at early times
\citep{Hicken07}.\footnote{Note that this value was derived by
  assuming a high value for the host-galaxy reddening of \gz, which
  might have been an overestimate.}  To reconcile this difference in
nickel mass,
it was suggested that perhaps \gz\ was a highly
asymmetric explosion, and due to the exact viewing angle more \nic\ was
inferred at early times than was actually present \citep{Maeda09}.
However, \dc\ appears to be spherically symmetric at early times
\citep{Tanaka09}. Furthermore, \citet{Maeda09} posit that dust might
have formed in the ejecta of \gz\ by this time, which would reprocess
much of the optical light to longer wavelengths.  However, we do not
see strong evidence for dust formation in \dc\ since the spectral
features all appear to evolve symmetrically.


To constrain the nickel mass produced by \dc\ from late-time
photometry, we write out the SN luminosity as a function of time 
\citep[e.g.,][]{Clocchiatti97,Maeda03, Maeda09},
\begin{equation}\label{eq:lum}
L(t) = M_{\rm Ni} \left(\varepsilon_\gamma (1 - e^{-\tau(t)})+f_{{\rm e}^+}\varepsilon_{{\rm e}^+} \right) e^{-t/111.3},
\end{equation}
where $t$ is the number of days from the explosion date, $M_{\rm Ni}$
is the initial amount of \nic\ synthesised, $\varepsilon_\gamma = 6.8
\times 10^9 {\rm ~erg~s}^{-1} {\rm ~g}^{-1}$, $\tau(t)$ is the
time-dependent optical depth to $\gamma$-rays, $f_{{\rm e}^+}$ is the
fraction of positrons trapped in the ejecta, $\varepsilon_{{\rm e^+}}
= 2.4 \times 10^8 {\rm ~erg~s}^{-1} {\rm ~g}^{-1}$, and 111.3~days is
the $e$-folding time of $^{56}$Co decay. Evaluating $\tau(t)$ requires
defining a model with estimates for the kinetic energy of the
explosion and the total mass of the white dwarf.

We can place constraints on the \nic\ mass by only considering the
luminosity powered by the trapped positrons; to do this we set
$\tau(t)=0$ in Eqn.~\eqref{eq:lum}. Using only the optical
contribution to the bolometric luminosity over the range
4000--8000~\AA\ from our Keck data taken 281~days past maximum (which
is 304~days after explosion), we calculate the \nic\ mass for a number
of reddening and $f_{{\rm e}^+}$ values which can be found in Table
\ref{t:late_time_ni}.
Our calculation of the bolometric luminosity does not include a
bolometric correction for any possible contribution to the bolometric
luminosity in the NIR.
For our nominal values of $E(B - V)_{\rm host} =0.1$~mag, $R_{V} =
3.1$, and $f_{{\rm e}^+} = 1$, we calculate a \nic\ mass of $1.4 \pm
0.3$~\msun, consistent with the \nic\ mass derived from Arnett's law
and our data at \bmax\ (\S\ref{sss:arnett}).

\begin{table}
\begin{center}
\caption{Ni mass estimates from late-time photometry
\label{t:late_time_ni}}
\begin{tabular}{c c c c | c}
\hline
Days past &
$f_{{\rm e}^+}$ &
$E(B-V)_{\rm host}$   &
$L_\textrm{bol}^\textrm{a} $  &
\nic \\
Explosion& & (mag)& ($10^{40}$~\ergps) & (\msun) \\
\hline
304 & 0.75 & $ 0 $ & $3.1 \pm 0.7$ & $1.4 \pm 0.3 $ \\
304 &1 &  $ 0 $ & $3.1 \pm 0.7$  &  $1.0 \pm 0.2 $ \\
304 &0.75 & $ 0.1 $ & $4.2 \pm 0.8$ &  $1.8 \pm 0.3$ \\
304 &1 &  $ 0.1 $ & $4.2 \pm 0.8$  & $1.4 \pm 0.3$ \\
304 &0.75 & $ 0.3 $ & $7.9 \pm 1.5$ &  $3.4 \pm 0.6$ \\
304 &1 &  $ 0.3 $ & $7.9 \pm 1.5$  & $2.5 \pm 0.6$ \\
\hline
426 & 0.75 & $ 0 $ & $0.26 \pm 0.05$ & $0.33 \pm 0.06 $ \\
426 &1 &  $ 0 $ & $0.26 \pm 0.05$  &  $0.25 \pm 0.05 $ \\
426 &0.75 & $ 0.1 $ & $0.35 \pm 0.07$ &  $0.45 \pm 0.08$ \\
426 &1 &  $ 0.1 $ & $0.35 \pm 0.07$  & $0.34 \pm 0.06$ \\
426 &0.75 & $ 0.3 $ & $0.64 \pm 0.12$ &  $0.82 \pm 0.15$ \\
426 &1 &  $ 0.3 $ & $0.64 \pm 0.12$ & $0.62 \pm 0.12$ \\

\hline
\end{tabular}
\end{center}
$^\textrm{a}$Bolometric luminosity only contains contribution over
the range 4000--8800~\AA.
\end{table}

We also determine the \nic\ mass using our epoch of photometry from
403~days past maximum (which is 426~days after explosion). However,
without a spectrum to model the spectral energy distribution at this
phase, we had to use our last spectrum taken 281~days past
\bmax, which could lead to an inaccurate measure of the bolometric
luminosity.  Furthermore, the detection of the SN was marginal in all
bands. Using our nominal values of $E(B - V)_{\rm host} =0.1$ mag,
$R_{V} = 3.1$, and $f_{{\rm e}^+} = 1$, we calculate a \nic\ mass of
$0.34 \pm 0.06$ \msun, clearly much lower than previous estimates of
the \nic\ mass. We summarise \nic\ masses for a range of parameters in
Table \ref{t:late_time_ni}.

The drop in derived \nic\ mass and expected luminosity could indicate
the formation of dust between $t = 281$ and 426 days past maximum
light, although this cannot be substantiated without IR data or a
relatively high-S/N spectrum.  The unexpected drop could also be an
indication that the positron trapping fraction at this phase is
significantly lower than the nominal value of $f_{{\rm e}^+} = 1$.  To
produce the \about1.4~\msun\ of \nic\ found using our photometry from
281~days past maximum, we would need $f_{{\rm e}^+} \approx
0.25$. \dc\ could have undergone an ``IR catastrophe'' which would
cause most of the emission to escape at IR wavelengths \citep[as
  opposed to the optical;][]{Axelrod88} as the temperature of the SN
cools to \about1000~K.  This would require using a bolometric
correction to our derived optical bolometric
luminosity. Alternatively, the low derived \nic\ mass could suggest
that the late-time light curve of \dc\ is not powered by the decay of
$^{56}$Co. This conclusion, however, is at odds with the \nic\ mass
derived from our epoch of photometry from $t = 281$~days past maximum
light. Without near-IR data or a spectrum at this epoch, it is
difficult to reconcile these differences. We again caution that the
spectral energy distribution is ill constrained at this phase.



\subsubsection{WD Masses and Kinetic Energies}\label{sss:energy}

Using the technique developed by \citet{Howell06}, we attempt to
combine our observations of \dc\ with properties of WDs and SNe~Ia in
order to calculate the mass of the progenitor WD as well as the amount
of \nic\ produced in the explosion.  We note that this method is
independent of the one used in the previous section which utilitized
the late-time photometry of \dc. As will be mentioned below, there 
are a few assumptions and simplifications that go into this analysis,
and thus the final values should probably not be taken literally.
However, the sense of the relationships and trends these calculations
indicate should be reliable and will show that \dc\ likely had a SC~WD
progenitor.

The ejecta of SNe~Ia derive their kinetic energy ($E_K$) from the
energy released during explosive nucleosynthesis ($E_n$), but the
ejecta must first overcome the binding energy ($E_b$) of the WD
progenitor \citep{Branch92}.  Therefore, we can write
\begin{equation}
E_K = \frac{M_\textrm{WD} v^2}{2} = E_n-E_b,
\end{equation}
where $M_\textrm{WD}$ is the mass of the WD just before explosion and
$v$ is a representative velocity of the ejecta.  A single value for
velocity only relates to the kinetic energy of the SN averaged over
the entire ejecta, and thus there has been disagreement as to which
actual value should be used for a given SN.  Previous studies have
used the velocity of the \ion{Si}{II} $\lambda$6355 feature at 10 and 
40~days past maximum \citep{Benetti05,Howell06}.

Since about 70\% of IGEs produced in SNe~Ia are in the form of
\nic\ \citep{Nomoto84,Khokhlov93}, the total mass of the WD can be
related to the nickel mass by
\begin{equation}
M_\textrm{Ni} = 0.7 M_\textrm{WD} f_\textrm{IGE},
\end{equation}
where $f_\textrm{IGE}$ is the fractional composition of IGEs in the SN
ejecta.  However, the exact amount of IGEs that are made up of
\nic\ may not be the same for all SNe~Ia and likely dependent on
other, external factors such as metallicity \citep[e.g.,][and
  references therein]{Howell09}.

Furthermore, \citet{Branch92} found that fusing equal parts carbon and
oxygen all the way to iron yields $\varepsilon_\textrm{Fe} =
1.55\times10^{51}$~erg~\msun$^{-1}$, and burning the same material
only up to silicon liberates about 76\% of that energy.  Thus we can
write
\begin{equation}
E_n = \varepsilon_\textrm{Fe} M_\textrm{WD} \left(f_\textrm{IGE} + 0.76
  f_\textrm{IME}\right),
\end{equation}
where $f_\textrm{IME}$ is the fractional composition of IMEs in the ejecta.

Finally, we define the fractional composition of unburned carbon
in the SN ejecta as $f_\textrm{C}$.\footnote{This parameter does not
appear in the equation for $E_n$; it represents the amount of 
{\it unburned} material, and thus does not contribute to the energy
released by nucleosynthesis.} Based on our definitions we can write 
\begin{equation}
1 = f_\textrm{C} + f_\textrm{IME} + f_\textrm{IGE}.
\end{equation}
Putting all of this together, we relate various properties of the
progenitor WD to properties of the SN ejecta it created using
\begin{equation}
v^2 \approx 2\varepsilon_\textrm{Fe} \left(\frac{0.34M_\textrm{Ni}}{M_\textrm{WD}} -
  0.76f_\textrm{C} + 0.76\right) - \frac{2E_b}{M_\textrm{WD}}.
\end{equation}
However, given the assumptions and simplifications that went into this
``equation'', we stress that it should be used more as an illustration
of how the different parameters relate to each other and less as a
tool to actually calculate WD and \nic\ masses.

Values of $E_b$ range from about $0.5\times10^{51}$~erg for a WD with
a mass of 1.4~\msun\ to about $1.3\times10^{51}$~erg for a WD with a
mass of 2~\msun\ and a central density of $4\times10^9$ g cm$^{-3}$
\citep{Yoon05}.  We perform separate calculations for each pair of
these values of $E_b$ and $M_\textrm{WD}$. \citet{Tanaka09} measure
the \ion{Si}{II} $\lambda$6355 feature to be blueshifted by
7200~\kms\ in their spectrum of \dc\ obtained 6~days past maximum,
while we find the same feature to be blueshifted by
5000--6000~\kms\ in our spectra taken 35, 52, and 64~days past
maximum.  Since $v$ has been evaluated at 10 and 40~days past maximum
for similar analyses previously \citep{Benetti05,Howell06}, we will
use a range of velocities from 7000~\kms\ to 5500~\kms\ for our
analysis. Finally, since there is clear indication of unburned
material in the ejecta of \dc\ (see \S\ref{sss:carbon}), we know that
$f_\textrm{C}$ must be nonzero.  However, based on models of SN~Ia
ejecta \citep[e.g.,][]{Nomoto84} and the relative scarcity of carbon
detections in SNe~Ia \citep{Marion06,Thomas07}, we will only use
values in the range 0.05--0.3 for $f_\textrm{C}$.

With such large uncertainties and so many parameters, we will clearly
calculate a huge range of \nic\ masses, and we realize that a few
assumptions and model-dependent values have come into the derivation.
However, this is still a useful line of reasoning in our attempt to
determine both the mass of the WD progenitor and the amount of
\nic\ synthesised by \dc.  We therefore forge ahead and attempt to
eliminate at least some of the most extreme situations.

Almost all of parameter space is ruled out when we use a 1.4~\msun\ WD
(and its associated binding energy), since we calculate negative
nickel masses for these cases.  The largest \nic\ mass we can
reasonably derive for a 1.4~\msun\ WD (using $f_\textrm{C} = 0.3$ and
$v = 7000$~\kms) is 0.06~\msun, nearly an order of magnitude smaller
than the amount of nickel produced by a normal SN~Ia
\citep[e.g.,][]{Nomoto84,Kasen08}.  We have also shown that \dc\ is
much more luminous than the average SN~Ia. Therefore, if it did have a
1.4~\msun\ progenitor, then we would need to invoke a large, hitherto
unknown energy source to power its light curve, which seems highly
unlikely.

The situation changes when we use a 2~\msun\ WD (with a central
density of $4\times10^9$ g cm$^{-3}$), put forth as a possible
SC~SN~Ia progenitor by \citet{Yoon05}.  Again, a significant part of
parameter space is eliminated, since at low values of $f_\textrm{C}$ we
calculate negative (or extremely small) nickel masses. 
Assuming that $f_\textrm{IME}$ must be $\gtrsim0.1$ \citep[given that
  we detect IMEs in the spectra for months after maximum; see,
  e.g.,][]{Nomoto84}, we calculate nickel masses as high as
\about1~\msun.
In order to match our nominal \nic\ mass range of 1.4--1.7~\msun\ 
{\it and} satisfy our constraints on the fractional composition 
of elements mentioned previously, we must resort to using a WD
progenitor more massive than 2~\msun. 


Again, the final values calculated above are almost certainly not the
true answer.  However, the range of plausible values presented, as
well as the relationships between the parameters of the SN and its WD
progenitor, yet again indicate that the progenitor of \dc\ was likely
a SC~WD.

The various analyses presented above seem to strongly favor the
conclusion that the progenitor of \dc\ was a SC~WD with a mass of
probably greater than \about2~\msun.  These analyses also indicate
that \dc\ most likely produced 1.4--1.7~\msun\ of \nic\ (assuming our
nominal value for the host-galaxy reddening and our peak bolometric
luminosity). More than 1~\msun\ of \nic\ was almost certainly created
by \dc, and the actual value {\it could} be as high as
\about3.3~\msun.  This matches well with the conclusions of
\citet{Yamanaka09}, who calculate a similar range of nickel masses for
\dc\ (1.2--1.8~\msun).  Part of the difference can be accounted for by
the fact that the two studies use different values of host-galaxy
reddening (causing the derived bolometric luminosities to differ
somewhat; see \S\ref{ss:lum}).  However, the majority of the
difference in derived \nic\ masses comes from the assumed rise time of
\dc.  \citet{Yamanaka09} adopt a rise time of 20~days based on
comparisons to typical SNe~Ia and \gz, while we use a rise time of
23~days based on our pre-maximum photometry. The longer rise time used
in our study leads to a larger derived nickel mass for \dc.

\fg\ had a derived progenitor mass of \about2.1~\msun\ and produced
\about1.3~\msun\ of \nic\ \citep{Howell06}, and these values are quite
similar to those we calculate for \dc.  \gz\ was estimated to have
produced 1--1.2~\msun\ of \nic, which is on the low end of our range
for \dc, and it was also claimed to have a SC~WD progenitor
\citep[though no attempt was made to further constrain the progenitor
  mass;][]{Hicken07}.

\subsection{Comparison to Theoretical Models}\label{ss:models}

Any theoretical model which is postulated to explain \dc, with or
without a SC WD, {\it must} be able to reproduce the observed
peculiarities for which we have very tight constraints: ({\it a}) high
luminosity even when assuming no host-galaxy reddening, ({\it b})
relatively long light-curve rise time, ({\it c}) relatively slow
photometric decline and late-time spectroscopic evolution, ({\it d})
the presence of carbon in spectra near maximum brightness, ({\it e})
the presence of silicon in spectra as late as a few months past
maximum brightness, ({\it f}) IGEs dominating the spectra at late
times, and ({\it g}) mostly spherically symmetric ejecta near maximum
with possible clumpy layers of IMEs \citep{Tanaka09}.

Below we consider both SC and non-SC models, all of which involve the
thermonuclear explosion of a WD.  However, the similarities between
\dc\ and SN~2002cx and its brethren (see \S\ref{sss:cx}) may argue
that the progenitor and explosion scenario for these objects are all
linked.  Could \dc\ be a SN~2002cx-like object but with a much higher
luminosity?  Perhaps, although attempts to explain the origins of even
the best-observed SN~2002cx-like objects are still ongoing
\citep[e.g.,][]{Jha06,Phillips07,Valenti09,Li10:02cx}.

\subsubsection{SC Models}

Two- and three-dimensional models of differentially rotating massive
WDs have been presented in the literature, and calculations show that
SC~WDs with masses as large as 2~\msun\ are possible by accretion from
a nondegenerate companion, while masses up to \about1.5~\msun\ are
possible from a double-degenerate merger
\citep[][respectively]{Yoon05,Yoon07}.  More recent studies of
double-degenerate mergers have even shown possible SN~Ia progenitors
with total masses approaching 2.4~\msun\ \citep{Greggio10}.  This is
encouraging for the case of \dc, since our energetics arguments imply
that its progenitor is likely greater than \about2 \msun. However,
calculations by \citet{Piro08} suggest that differential rotation is
unlikely for WDs accreting from a nondegenerate companion, and thus
their masses cannot exceed the Chandrasekhar mass by more than a few
percent.


\citet{Yoon05} also consider how much \nic\ would be produced by a
SN~Ia whose progenitor is \about2~\msun, calculating values of
0.4--1.3~\msun.  This large range does encompass the lower end of our
range of \nic\ for \dc.  However, only specific kinetic energies
(i.e., kinetic energies per unit mass) which are lower than what we
find for \dc\ by a factor of \about2 were used \citep{Yoon05}.
Adopting lower specific kinetic energies will tend to decrease the
nickel yield for a given WD mass, according to their models.  It is
possible that increasing the specific kinetic energies used in the
models of \citet{Yoon05} to values that better match \dc\ will in fact
reproduce our derived nickel mass, but this may require pushing the
models into regimes where they are no longer valid.

\citet{Pfannes09} use aspects of the WD models of \citet{Yoon05} and
simulate prompt detonations of massive, rapidly rotating WDs.  None of
their models appears to reproduce the fact that we detect unburned fuel
(i.e., carbon) and IMEs near maximum brightness at low
velocities, and subsequently 
IMEs and IGEs at later times (also at low velocities).
Even more at odds with \dc, \citet{Pfannes09} predict
that the IMEs and unburned material should have similar spatial
distributions in the ejecta (specifically, in a torus in the
equatorial plane), but this seems unlikely given the strong  
line polarizations seen by \citet{Tanaka09} in \ion{Si}{II} and
\ion{Ca}{II} features and the lack thereof in two different \ion{C}{II} 
features.  Thus, we conclude that the models of \citet{Pfannes09} do
not seem to reproduce the observations of \dc.

Models of SC~WDs and their evolution in close binary systems with
nondegenerate companions can also be found in the literature
\citep{Chen09}. These evolutionary calculations include the effects of
varying the orbital period of the binary systems, the metallicity and
mass-transfer rate of the binary companion, and (most importantly) the
mass of the WD.  \citet{Chen09} find that WDs with initial masses of
\about1.2~\msun, under the right mass-transfer conditions, can accrete
up to masses of about 1.4--1.8~\msun\ before exploding as a SN~Ia
(though they find that most of these WDs explode with masses not much
above 1.4~\msun).  Thus it seems somewhat unlikely that these models
can explain the progenitor we suggest for \dc; even if the
occasional WD {\it can} accrete up to 1.8~\msun, this is at the lowest
end of our range of WD progenitor masses.

Spherically symmetric, one-dimensional radiation transport
calculations for normal and SC~WDs have been carried out by
\citet{Maeda09sc}.  They find that based on the light-curve shapes,
photospheric velocities, peak bolometric luminosities, and peak
effective temperatures, \gz\ likely came from a
SC~WD\footnote{\citet{Maeda09sc} claim that most of the emission
  likely shifted to the near-IR and mid-IR at late times in order to
  explain the relatively faint late-time observations of
  \citet{Maeda09}.}  while \fg\ did not.  Since the observables of
\fg\ used for their analysis have very similar values to those of \dc,
it would seem that their models imply that \dc\ is also not a
SC~SN~Ia.  \citet{Maeda09sc} point out that \fg\ (and thus probably
\dc\ as well) could be a SC~SN~Ia if the progenitor star were highly
aspherical, but again this seems unlikely for \dc\ from the
spectropolarimetric data \citep{Tanaka09}.

The primary independent variable used in the calculations of
\citet{Maeda09sc} is $t_{+1/2}$, the number of days after maximum
bolometric luminosity when the bolometric luminosity has decreased to
half its maximum value \citep{Contardo00}.  For \gz\ they measure
$t_{+1/2}=18$~days from the publicly available photometry, but for
\fg\ they convert the stretch value published by \citet{Howell06} to
$\Delta m_{15}$ and then to a $t_{+1/2}$ value of 13.5~days
\citep[adopting an empirical linear fit to $\Delta m_{15}$ and
  $t_{+1/2}$ values from][]{Contardo00}. Using their conversions
between $\Delta m_{15}$ and $t_{+1/2}$, we calculate $t_{+1/2} \approx
14$~days \citep[which, unsurprisingly, is nearly the same as the value
  calculated for \fg\ by][]{Maeda09sc}.  However, when we measure
$t_{+1/2}$ directly from our light curve, we get a minimum value of
about 20.5~days (when we use our maximum plausible host-galaxy
reddening), which is nearly 50\% larger!  Furthermore, if one converts
the $\Delta m_{15}$ value of \gz\ to $t_{+1/2}$ using the linear fit,
one again finds $t_{+1/2} \approx 14$~days, suggesting that the linear
conversion derived by \citet{Maeda09sc} is not accurate for such low
values of $\Delta m_{15}$.  This is not wholly unexpected since the
$\Delta m_{15}$ values of SNe~2006gz, 2003fg, and 2009dc (or
$t_{+1/2}$ as measured from light curves) are all below (or above) the
values used to derive the linear fit \citep{Contardo00}.  In addition,
the conversion from stretch to $\Delta m_{15}$ used by
\citet{Maeda09sc} for \fg\ did not take into account the fact that the
definition of stretch has evolved as new SN~Ia light-curve templates
have been constructed \citep[e.g.,][]{Conley08}.

We can now compare \dc\ to the analysis of \citet{Maeda09sc} using the
actual values of $t_{+1/2}\approx 20.5$--$22.8$~days. Given the large
bolometric luminosity and $t_{+1/2}$ value, as well as the extremely
low photospheric velocity of \dc, none of their models (using normal
or SC WDs) appears to be viable. Their ``normal WD'' model with
$M_\textrm{WD}=1.39$~\msun\ and $M_\textrm{Ni}=0.6$~\msun\ accounts
for the low velocity with large $t_{+1/2}$, but it underpredicts
$L_\textrm{bol}$ by a factor of a few.  Some of the SC~WD models
($M_\textrm{WD}\ge2.0$~\msun\ and $M_\textrm{Ni}=1.0$ \msun) of
\citet{Maeda09sc} can almost account for the large values of
$L_\textrm{bol}$ and $t_{+1/2}$ seen in \dc, but they then overpredict
the photospheric velocity of \dc\ by a factor of 2 or so (with the
models having the highest mass WDs coming the closest to matching the
observed velocities).

While no single model of \citet{Maeda09sc} clearly reproduces our
observations of \dc, its properties seem to be on the outskirts of the
parameter space explored by their analysis.  This hints at the
possibility that more extreme WD and \nic\ masses may be required to
match the observational data of \dc. In addition, it is possible that
{\it multi-dimensional} analyses are required to truly capture the
underlying physics of a SC~WD explosion.

Furthermore, a few synthetic late-time spectra of SC~SNe~Ia have
been presented in the literature \citep[e.g.,][]{Maeda09}.  Their
models take as inputs a WD progenitor mass, a WD central density, and
mass fractions of burning products, and then use a one-dimensional
Monte Carlo 
radiation transport code, along with ionization/recombination
equilibrium and rate equations, to calculate synthetic light curves and
spectra.  Both of their SC~SN~Ia models roughly match our \dc\
photometry.
In addition, their late-time spectra are
similar to our day 281 spectrum of \dc.  \citet{Maeda09} note that as
they increase their models' progenitor mass, the
[\ion{Fe}{II}]/[\ion{Ca}{II}] feature near 7200~\AA\ becomes stronger
relative to the blends of [\ion{Fe}{II}] and [\ion{Fe}{III}] near
3800--5500~\AA.  This is seen in \dc\ (as compared to SNe~Ia with more
normal peak luminosity), and it further supports our finding that the
progenitor of \dc\ was a SC~WD.  However, it should be noted that the
SC~SN~Ia models of \citet{Maeda09} only use WDs with masses of 2 and
3~\msun, and \nic\ masses of only 1~\msun, which is on the low end of
our range of calculated values for the \nic\ yield of \dc.

\subsubsection{Non-SC Models}

A number of models which employ Chandrasekhar-mass WDs as the
progenitors of super-luminous SNe~Ia (ones that we would consider
possibly SC~SNe~Ia) have been proposed
\citep{Hillebrandt07,Sim07,Kasen08,Kasen09}. These models usually
invoke the off-centre ignition of a normal WD progenitor which leads
to nuclear burning (and thus \nic\ production) that is peaked away
from the centre of the WD.  This off-centre nickel blob would increase
the observed luminosity if the blob were offset from the centre of the
WD toward the observer, with the maximum effect occurring when it is
offset directly along the line of sight. These viewing angles also
lead to the fastest light-curve rise times in such simulations
\citep{Sim07,Kasen08}.

Once again, we can compare our observed values for \dc\ to the models.
Since the nickel blob is offset from the centre, asphericity is
introduced into the explosion by construction \citep{Kasen08}.  If
there is in fact a blob of \nic\ in \dc\ and it is offset from the
progenitor's centre directly along our line of sight (thus maximizing
the measured luminosity), then perhaps there is still azimuthal
symmetry in the explosion.  This may account for the low levels of
continuum polarization measured by \citet{Tanaka09}, as well as the
higher levels of IME line polarization, but it seems tenuous at best.

Observing along the axis of the offset blob of \nic,
\citet{Hillebrandt07} produce light curves that peak at a bolometric
magnitude of $M_\textrm{bol}\approx -19.9$~mag, \citet{Kasen08} show
light curves that get as bright as $M_B \approx -20$~mag, and
\citet{Kasen09} claim models with luminosities as high as
$2.1\times10^{43}$~\ergps.  These brightest magnitudes and
luminosities that can be obtained by off-centre explosions are nearly
equal to our {\it lower limits} for \dc\ (assuming no host-galaxy
reddening), and are likely below the true values. Furthermore, we note
that the maximum amount of \nic\ obtained from these explosions is
0.9--1.1~\msun\ \citep{Hillebrandt07,Kasen09}, which is once again at
the lowest end of our range of calculated values for \dc.

Finally, as mentioned above, the viewing angles that maximize the
observed peak magnitudes also {\it minimize} the rise time.  We find a
relatively long rise time of $23 \pm 2$~days for \dc, and our first
detection of the SN implies that its rise time {\it must be}
$>$21~days.  This is significantly longer than the rise times for
these viewing angles as derived from the models \citep[\about12~days
  and \about18~days;][respectively]{Hillebrandt07,Kasen08}.  Thus, it
seems that none of these models which include a Chandrasekhar-mass WD
is viable for \dc.

It should be mentioned, however, that these models mostly assume
expansion velocities of ``normal'' SNe~Ia \citep[e.g.][]{Sim07} and we
have shown that the expansion velocity of \dc\ is significantly lower.
This is evident in the synthetic spectra derived from these models;
from all viewing angles, they resemble early-time normal SN~Ia
spectra much more than the spectra of \dc\ near maximum brightness
\citep{Kasen08}. Thus, perhaps it is not too surprising that these
specific examples of non-SC models do not match the observations of
\dc.


\section{Conclusions}\label{s:conclusions}

In this paper we have presented and analysed optical photometry and
spectra of \dc\ and \snif, both of which are possibly SC~SNe~Ia.  Our
photometric and spectral data on \dc\ constitute one of the richest
datasets ever published on a SC~SN~Ia candidate.  Our well-sampled
light curve follows \dc\ from about 1 week before maximum brightness
until about 5~months past maximum, and shows that \dc\ is one of the
slowest photometrically evolving SNe~Ia ever observed.  We derive a
rise time of 23~days and $\Delta m_{15} (B)=0.72$~mag, which are two
of the most extreme values for these parameters ever seen in a SN~Ia.
Assuming no host-galaxy reddening, we derive a peak bolometric
luminosity of about $2.4\times10^{43}$~\ergps, though this is almost
certainly an underestimate since we observe strong evidence for at
lease {\it some} host reddening.  Using our nonzero values for
$E(\bv)_\textrm{host}$, the peak bolometric luminosity increases by
about 40\%--200\%.

Spectroscopically, \dc\ also evolves relatively slowly.  Strong
\ion{C}{II} absorption features (which are rarely observed in SNe~Ia)
are seen in the spectra near maximum brightness, implying a
significant amount of unburned fuel from the progenitor WD in the
outer layers of the SN ejecta.  \ion{Si}{II} absorption also appears
in our spectra of \dc\ and remains visible even 2~months past maximum.
Our post-maximum spectra are dominated by a forest of IGE features
and, interestingly, resemble spectra of the peculiar SN~Ia 2002cx.
Finally, the spectra of \dc\ all show very low expansion velocities at
all layers (i.e., unburned carbon, IMEs, and IGEs) as compared to
other SNe~Ia. This may be explained by a massive WD progenitor which
consequently has a large binding energy.  Even though the expansion
velocities are small, we see no strong evidence in \dc\ for a velocity
``plateau'' near maximum light like the one seen in \snif\
\citep{Scalzo10}. 

Using various luminosity and energy arguments, we calculate that the
progenitor of \dc\ is possibly a SC~WD with a mass greater than
\about2~\msun, and that at least \about1~\msun\ of \nic\ was likely
formed in the explosion (though the most probable value is in the
range 1.4--1.7~\msun).  These values are larger than (or about as
large as) those calculated for any other SN~Ia ever observed.  We
propose that the host galaxy of \dc\ underwent a gravitational
interaction with a nearby galaxy (UGC~10063) in the relatively recent
past, and that this could have induced a sudden burst of star
formation which may have given rise to the progenitor of \dc\ and
turned UGC~10063 into the ``post-starburst'' galaxy that we observe
today.  We also compare our observed quantities for \dc\ to
theoretical models, and while no model seems to match or explain every
aspect of \dc, simulations show that SC~WDs with masses near what we
calculate for the progenitor of \dc\ can possibly form, likely from
the merger of two WDs.  Furthermore, models of extremely luminous
SNe~Ia which employ a Chandrasekhar-mass WD progenitor cannot explain
our observations of \dc.

Thus, taking all of these extreme values into account, we conclude
that \dc\ is very likely a SC~SN~Ia.  As mentioned previously, many of
the observed peculiarities of \dc\ are also seen in \fg\ and \snif.
Therefore, we concur with \citet{Howell06} and \citet{Scalzo10} that
both \fg\ and \snif\ (respectively) are also probably SC~SNe Ia.
However, given their fairly normal expansion velocities and relative
weakness (or even absence) of \ion{C}{II} features near maximum
brightness, it seems that \gz\ and SN~2004gu are less likely to be
SC~SNe~Ia.


New large transient searches such as Pan-STARRS \citep{Kaiser02} and
the Palomar Transient Factory \citep[][]{Rau09,Law09} will probably
find many SC or other super-luminous SNe~Ia in the near future.  Since
it seems that they cannot be standardized in the same way as most
SNe~Ia, they will need to be handled separately or ignored in
cosmological surveys which will use large numbers of SNe~Ia.  However,
the simulations of \citet{Chen09} show that donor stars with lower
metallicities (e.g., Population II stars) are less likely to form WDs
with masses greater than 1.7~\msun\ than higher metallicity
stars. Thus, it is possible that contamination levels from SC~SNe~Ia,
which are already rare at low redshifts (i.e., average metallicity),
may be relatively small in medium or high-redshift surveys.

\section*{Acknowledgments}

We thank P.~F.~Hopkins, K.~M.~Sandstrom, and K.~L.~Shapiro for useful
discussions regarding host galaxies.  We are especially grateful to
D.~A.~Howell for the near-maximum spectrum of \fg\ (and, as the
referee, for many useful comments), K.~Maeda for the late-time
spectrum of \gz, R.~A.~Scalzo and P.~E.~Nugent for information
regarding \snif, and M. Yamanaka for early-time comparison photometry
of \dc.  We also thank I.~Arcavi, S.~B.~Cenko, J.~Choi, B.~E.~Cobb,
R.~J.~Foley, C.~V.~Griffith, M.~T.~Kandrashoff, M.~Kislak,
I.~K.~W.~Kleiser, J.~Leja, M.~Modjaz, A.~J.~L.~Morton, J.~Rex,
T.~N.~Steele, P.~Thrasher, and X.~Wang for their assistance with some
of the observations and data reduction, as well as J.~Kong, N.~Lee, and
E.~Miller for their help in improving and maintaining the SNDB.  We
are grateful to the staffs at the Lick and Keck Observatories for
their support.  Some of the data presented herein were obtained at the
W.~M.~Keck Observatory, which is operated as a scientific partnership
among the California Institute of Technology, the University of
California, and the National Aeronautics and Space Administration
(NASA); the observatory was made possible by the generous financial
support of the W.~M.~Keck Foundation. The authors wish to recognize
and acknowledge the very significant cultural role and reverence that
the summit of Mauna Kea has always had within the indigenous Hawaiian
community; we are most fortunate to have the opportunity to conduct
observations from this mountain.  This research has made use of the
NASA/IPAC Extragalactic Database (NED) which is operated by the Jet
Propulsion Laboratory, California Institute of Technology, under
contract with NASA.  A.V.F.'s group is supported by the NSF grant
AST--0908886, DOE grants DE--FC02--06ER41453 (SciDAC) and
DE--FG02--08ER41563, NASA/{\it Swift} grant NNX09AL08G, and the
TABASGO Foundation.  KAIT and its ongoing operation were made possible
by donations from Sun Microsystems, Inc., the Hewlett-Packard Company,
AutoScope Corporation, Lick Observatory, the NSF, the University of
California, the Sylvia \& Jim Katzman Foundation, and the TABASGO
Foundation. J.M.S. is grateful to Marc J. Staley for a Graduate
Fellowship.



\bibliographystyle{mn2e}

\bibliography{astro_refs.bib}

\label{lastpage}

\end{document}